\documentclass[11pt]{article}

\usepackage{geometry} 
\usepackage{verbatim}
\usepackage{caption}
\usepackage{subcaption}
\usepackage{authblk}
\usepackage{epsfig,latexsym}
\usepackage{amsmath,amssymb,amsthm,graphicx,verbatim,bm,fancyhdr,bbm,enumerate,color}

\DeclareMathAlphabet{\mathpzc}{OT1}{pzc}{m}{it}  

\theoremstyle{definition}
\theoremstyle{plain}
\newtheorem{theorem}{Theorem}[section]
\newtheorem{proposition}[theorem]{Proposition}

\newtheorem{corollary}[theorem]{Corollary}

\newtheorem{definition}{Definition}
\renewenvironment{proof}{{\bfseries Proof.}}{\hfill $\blacksquare$}

\theoremstyle{remark}

\renewcommand{\P}{\mathbb{P}}

\newcommand{\E}{\mathbb{E}}

\newcommand{\R}{\mathbb{R}}

\newcommand{\cE}[2]{\mathbb{E}\left(#1\,\left\vert\vphantom{#1#2}\right. #2\right)}
\newcommand{\cP}[2]{\mathbb{P}\left(#1\,\left\vert\vphantom{#1#2}\right. #2\right)}
\newcommand{\cEv}[2]{\left(#1\,\left\vert\vphantom{#1#2}\right. #2\right)}

\newcommand{\ve}{\varepsilon}

\newcommand{\set}[1]{\left\{#1\right\}} 
\newcommand{\ob}[1]{\left(#1\right)} 
\newcommand{\cb}[1]{\left[#1\right]} 

\begin{document}

\title{Spatial measures of genetic heterogeneity during carcinogenesis}
\author[1]{Kathleen Storey}
\author[2]{Marc D. Ryser}
\author[3]{Kevin Leder}
\author[4]{Jasmine Foo}
\affil[1]{{\small School of Mathematics, University of Minnesota, Minneapolis, MN, USA}}
\affil[2]{{\small Department of Mathematics, Duke University, Durham, NC, USA}}
\affil[3]{{\small Industrial and Systems Engineering, Minneapolis, MN, USA, lede0024@umn.edu}}
\affil[4]{{\small School of Mathematics, Minneapolis, MN, USA, jyfoo@umn.edu}}
\date{\today}
\maketitle
 
 \begin{abstract}
 
In this work we explore the temporal dynamics of spatial heterogeneity during the process of tumorigenesis from healthy tissue. We utilize a spatial stochastic process model of mutation accumulation and clonal expansion in a structured tissue to describe this process. Under a two-step tumorigenesis model, we first derive estimates of a non-spatial measure of diversity: Simpson's Index, which is the probability that two individuals sampled at random from the population are identical, in the premalignant population.  We next analyze two new measures of spatial population heterogeneity. In particular we study the typical length scale of genetic heterogeneity during the carcinogenesis process and estimate the extent of a surrounding premalignant clone given a clinical observation of a premalignant point biopsy.  This evolutionary framework contributes to a growing literature focused on developing a better understanding of the spatial population dynamics of cancer initiation and progression. Although initially motivated by understanding questions in cancer, these results can be applied more generally to help understand the dynamics of heterogeneity and diversity in a variety of spatially structured, evolving populations.
 \end{abstract}
\section{Introduction}\label{intro}

Carcinogenesis, the transformation from healthy tissue to invasive cancer, is a lengthy and complex process driven by a variety of factors including hereditary predisposition \cite{rahman2014realizing}, exposure to environmental factors \cite{wild2013measuring} and a changing microenvironment in the affected organ \cite{gillies2015metabolism}. Irrespective of the driving factors, most cancers are characterized by the progressive accumulation of genetic alterations in a small group of founder cells. These alterations are either deleterious or neutral (passenger mutation), and some can confer a fitness advantage to the affected cell (driver mutation) by increasing the reproductive rate or inhibiting cell-regulatory mechanisms \cite{bozic2010accumulation}. These selective advantages in turn lead to clonal expansion of a mutant cell population, which provides a fertile backdrop for further genetic alterations. Importantly, the underlying tissue architecture strongly influences the spatial growth patterns of the premalignant lesions, leading to complex patterns of spatial heterogeneity caused by competing and overlapping clones of various sizes and genetic ancestries \cite{mcgranahan2015biological}.

The extent of spatial heterogeneity arising from this evolutionary process has been shown to correlate with clinical outcome. For example, genetic clonal diversity in premalignant tissue found in cases of Barrett's esophagus has been shown to predict progression to esophageal carcinoma  \cite{maley2006genetic}. However, the translation of heterogeneity into a patient-specific clinical progression markers remains challenging because multiple point biopsies per patient are needed to reliably ascertain the degree of heterogeneity. Thus, there is a critical need for quantitative tools that (i) inform optimal sampling strategies, (ii) infer the degree of heterogeneity in premalignant tissue based on sparse sample data, and (iii) predict the evolution of premalignant lesions and time scale of progression.

In this work we develop and analyze a cell-based stochastic model that describes the evolutionary process of cancer initiation in a spatially structured tissue. This model is a spatial version of a Moran population model, and has previously been analyzed in \cite{durrett2015spatial,DuFoLe15,foo2014multifocality}. Using a mesoscopic approximation of this model, we analyze two spatial measures of heterogeneity that are relevant for the clinical setting. First, we study the probability that two samples, taken a fixed distance apart from each other, are genetically identical. This corresponds to a spatial analog of Simpsons' Index, a traditionally non-spatial measure of diversity which is defined as the probability that two individuals sampled at random from a population are identical. This measure, taken as a function of the distance between samples, provides an estimate of the length scale of heterogeneity in the premalignant tissue. As a second measure of heterogeneity, we study the expected size of a premalignant lesion. This measure may be useful in scenarios where an isolated point biopsy indicates premalignant tissue without further information about the extent of the lesion. For both measures, we  determine how they evolve during the transformation from healthy tissue to onset of malignancy, and we characterize their dependence on cancer-specific parameters such as mutation rates and fitness advantages. Due to the general formulation of the model, these results provide a useful tool for studying how heterogeneity and the extent of premalignant lesions vary between different cancer types.  Although initially motivated by understanding questions in cancer, these results can be applied more generally to help understand the dynamics of heterogeneity and diversity in a variety of spatially structured, evolving populations.

 The influence of spatial structure on the diversity of evolving populations has previously been studied in the ecological literature. Within that context, R.H. Whittaker introduced the measures of $\alpha$-,$\beta$- and $\gamma$-diversity to denote the average species richness at the single habitat level ($\alpha$), the diversity between habitats ($\beta$), and total species richness ($\gamma$) \cite{whittaker1972evolution}. These measures are useful to quantify large scale organismal diversity in an ecological setting with spatial variation between well-defined habitats. However in the present work we are interested in developing new measures of diversity to explore more specifically the intrinsic length scales of genetic heterogeneity driven by clonal expansion dynamics in a spatially structured tissue population.

There have been other mathematical modeling efforts on the topic of heterogeneity during cancer initiation and expansion.  In particular, previous work by Iwasa and Michor explored the Simpson's Index in a Moran process of tumorigenesis \cite{IwMi2011}. This study focused on understanding the impact of neutral and advantageous mutations in a non-spatial, homogeneously mixed population setting. The work by Durrett and et. al. \cite{DuFoLeMaMi11} developed formulas for Simpson's Index and other heterogeneity measures in a multitype branching process model of cancer evolution. More recently, Dhawan and colleagues \cite{dhawan2015computational} developed a computational platform for the comparison of alternative spatial heterogeneity measures as potential biomarkers for tumor progression. Finally, within the broader context of spatial tumor growth, our work adds to a vast body of literature, e.g.\@ \cite{Ko06s, Nowak2003, WilBje72, ThLoStKo10, Ko13, durrett2015spatial}.  \\

The outline of this paper is as follows: In Section \ref{sec:model} we introduce a cell-based stochastic evolutionary model of spatial carcinogenesis, as well as a mesoscopic approximation to this model that was analyzed in \cite{DuFoLe15}. In Section \ref{sec:nonspatialHet} we first analyze the non-spatial Simpson's Index for this spatially-structured population. Then, in Section \ref{sec:SpatialHet} we formulate and analyze two clinically relevant spatial measures of heterogeneity and study their dependence on cancer-specific parameters. Finally, we summarize and discuss our findings in Section \ref{sec:sum}.

\section{Model}
\label{sec:model}
We introduce a spatial evolutionary model that describes the dynamic transition from physiological homeostasis to onset of invasive cancer. In between, the tissue undergoes a sequence of genetic changes that manifest themselves at the phenotypic level in the form of increased proliferation rates,  and hence a fitness advantage of mutant cells over normal cells. It is important to note that in many cancers, there is a succinct lack of a clearly defined genetic sequence \cite{sprouffske2011accurate}. On the other hand, the morphological changes from normal tissue to dysplasia, carcinoma in situ and invasive cancer is common in carcinomas, which account for over 80\% of all cancers. Therefore, one might prefer to interpret \emph{mutations} as phenotypic transitions rather than genetic aberrations. With this interpretation in mind, we are going to introduce a linear 3-stage model, where type 0 cells represent normal tissue, type 1 cells are pre-malignant (dysplasia/CIS) and type 2 cells are malignant cancer cells. Note that the model can be extended to a setting with more than two mutations, either to represent a more refined phenotypic progression, or to account for select cancer-specific genetic events.

To render this model spatial, we introduce a cell-based stochastic model on the integer lattice  $\mathbb{Z}^d\cap [-L/2,L/2]^d$, where $L>0$, and equip this domain with periodic boundary conditions. On this lattice we have three different types of cells, labeled as type 0, type 1 and type 2. For $i\in \{0,1,2\}$ a type $i$ cell reproduces at rate $(1+s)^i$, and when the cell reproduces it replaces one of its 2$d$ neighboring cells at random. In addition, we assume that for $i\in\{0,1\}$, a type $i$ cell mutates to type $i+1$ at rate $u_{i+1}$. Initially our entire lattice is occupied by type 0 cells which represent normal cells without any oncogenic mutations.  Tumor initiation is defined as the birth of the first type-$2$ cell that does not go extinct.  
In the biological application we are interested in (somatic cells in the body) $L$ is generally at least $10^6$ while $s$, $u_1$ and $u_2$ are quite small. Therefore we will, unless stated otherwise, restrict our analysis to the regime $L\ll1$, $u_1\ll1$, $u_2\ll1$, and $s\ll1$. Before we can discuss the specific conditions imposed on the model parameters, we need to review the dynamic properties of the model. \\

 In \cite{DuFoLe15} we established that the arrival of type-1 mutants that are successful (i.e. whose progeny does not go extinct) can be described as a Poisson process with rate $u_1s/(1+s)$. Here, $u_1$ is the mutation rate to type-1 and $s/(1+s)$ is the survival probability of each type-1 mutant.  We also characterized the radial expansion rate of type-1 families as a function of the selective advantage $s$ in each dimension.  In particular, it was established in \cite{BraGri80,BraGri81} that each successful type-1 family has an asymptotic shape $D$ which is a unit ball in an unknown norm, and grows linearly in time.  Let $e_1$ the unit vector in the first axis, and let $c_d(s)$ be the linear expansion rate of the radius of this ball:
$
D\cap\{ze_1:z\in\mathbb{R}\}=[-c_d(s),c_d(s)].$
Then, we established that as $s\to 0$,
$$
c_d(s) \sim \begin{cases} s & d=1 \\ \sqrt{4\pi s/\log(1/s)} & d=2 \\ \sqrt{4\beta_ds} & d \ge 3, \end{cases}
$$
where $\beta_d$ is the probability that two $d$ dimensional simple random walks started at $0$ and $e_1=(1,0,\ldots,0)$ never collide. \\

Based on this result, we then introduced a mesoscopic approximation to the model. Here, the growth of successful mutant families is deterministic while the arrival of these families follows a non-homogeneous Poisson process. To ensure that this mesocopic model accurately recapitulates the dynamics of the cell-based model, we will make the following assumptions on the relationships between parameters in the model: 
\begin{align}
&(A0) \qquad u_1 \ll 1/\ell(s)^{(d+2)/2}\\
&(A1) \qquad \left(\frac{c_d}{u_2s}\right)^{d/(d+1)}\ll N\nonumber\\
&(A2) \qquad (Nu_1s)^{d+1}(c_d^du_2s)^{-1}\to c\in [0,\infty)\\
&(A3) \qquad u_2 \ll 1/\ell(s)\nonumber
\end{align}
These assumptions generally hold for the parameter ranges appropriate for our biological application of carcinogenesis, see \cite{DuFoLe15} for a details. In addition, we will focus on dimensions $d=1$ and $d=2$ since most epithelial tissues can be viewed as one or two dimensional structures, e.g., the cells lining a mammary duct ($d=1$), the crypts in the colon ($d=2$), or the stratified squamous epithelia of bladder, the cervix and the skin ($d=2$).


In the simplified mesoscopic model we consider the cells to live on a spatial continuum $D=[-L,L]^d$. The state-space of the system is given by a set-valued function $\chi_t$, which characterizes the regions of $D$ occupied by type-1 cells at time $t$.  Mutations to type-1 cells occur as a Poisson process at rate $u_1s$ in the set $\chi_t^c=D\backslash\chi_t$, i.e. in regions where type-0 cells reside.  Each newly created type-1 mutation initiates an expanding ball whose radius grows linearly at rate $c_d$. Denoting the Euclidean ball by $B_{x,r}=\{y: ||y-x||\leq r\}$, then after $k$ mutations at the space-time points $\{(x_1,t_1),\ldots, (x_k,t_k)\}$, we have
$$
\chi_t=\bigcup_{i=1}^kB_{x_i,c_d(t-t_i)}.
$$
Thus, the state of the system at any time $t$ is the union of balls occupied by expanding mutant type-1 families.  In \cite{DuFoLe15} we proved that under assumption  (A2) we can neglect the possibility that a second mutation arises from a type-1 family that dies out eventually.  Therefore, we model successful type-2 mutations as Poisson arrivals into the space occupied by type-1 cells, $\chi_t$, with rate $u_2 s$.   Recall that in our two-step cancer initiation model, the type-2 mutant represents a malignant cancer cell.\\

We define the cancer initiation time $\sigma_2$ as the time when the first successful type 2 cell is born.   Then, $\sigma_2$ is a random variable with complimentary cumulative distribution function given by
$$
\P\left(\sigma_2>t\right)=\E\exp\left(-u_2s\int_0^t|\chi_t|dt\right),
$$
where $|\chi_t|$ is the area of type-1 cells at time $t$.
\section{Simpson's Index}
\label{sec:nonspatialHet}

Simpson's Index, a traditional non-spatial measure of heterogeneity, is defined as the probability that two individuals, sampled at random from a population, are genetically identical. More precisely, if there are $N$ types of individuals in a population, the Simpson's Index is defined as
 \begin{align}\label{SI}
R= \sum_{i=1}^{N} \ob{\frac{Y_{i}}{Y}}^2,
\end{align}
where $Y_i$ is the number of individuals of the $i$-th type, and $Y$ is the size of the entire population. Although this measure is usually used to characterize well-mixed populations, we investigate here how it evolves over time within the spatially structured population described by the mesoscopic spatial model from Section \ref{sec:model}. In the cancer setting, one question of interest is to determine the degree of heterogeneity of the premalignant cell population.  In our mesoscopic model, suppose there are $N_t$ type-1 clones present at time $t>0$.  We then extend definition (\ref{SI}) as the time-dependent quantity
\begin{align}\label{Rdef}
R(t)= \sum_{i=1}^{N_t} \ob{\frac{Y_{1,i}(t)}{Y_1(t)}}^2,
\end{align}
where $N_t$ is a Poisson random variable with parameter $N u_1 s t$, $Y_{1,i}$ denotes the volume of the type-1 subclone originating from the $i$th type 1 mutation, and $Y_1(t)$ is the total volume of all the type-1 families present at time $t$, i.e.\ $Y_1(t)=\sum_{k=1}^{N_t} Y_{1,i}(t)$. From conditions (A0)-(A3) and Theorem 4 of \cite{DuFoLe15} we know that overlaps between distinct type-1 clones occur with negligible probability by time $\sigma_2$; we therefore ignore this possibility in the computation of Simpson's Index.

Building on the theory of size-biased permutations, it is  possible to characterize the distribution of $R(t)$ as follows. 

\begin{proposition}
The conditional expectation of Simpson's Index for the spatial mesoscopic model is 
\begin{align}
\label{eq:Simpson2}
\E[R(t)|N_t=n]=n\E\left[\left(\frac{S_1}{S_n}\right)^2\right],
\end{align}
where $S_n:=B_1+\ldots + B_n$ with $B_i$ are i.i.d.\@ $Beta(\frac{1}{d},1)$ random variables.
\end{proposition}
The proof of this result is found in Appendix \ref{app:condexp}.
\begin{proposition}
The conditional variance of the Simpson's Index is bounded as follows
\begin{align}\label{upperboundSI}
\cb{\cE{ R(t)}{N_t=n}}^2 \leq  n\int_0^\infty \int_0^\infty  \ob{\frac{r}{x}}^3 \,  \nu_1(r)\, \nu_{n-1}(x-r)\,dx\,dr,
\end{align}
where $\nu_k$ is the probability density function of $S_k$ as defined above. 
\end{proposition}
The derivation of this bound is found in Appendix \ref{app:upboundvar}. Finally, the following result establishes the behavior of Simpson's Index for large $n$.
\begin{proposition}
\label{prop_Simpson_large_n}
Conditioned on $N_t=n$, $R(t)$ converges to zero in probability as $n\to\infty$.
\end{proposition}

This result tells us that as the number of clones increases, the probability of selecting two cells from the same clone goes to zero. 

\begin{figure}[htbp] 
   \centering
   \includegraphics[width=6in]{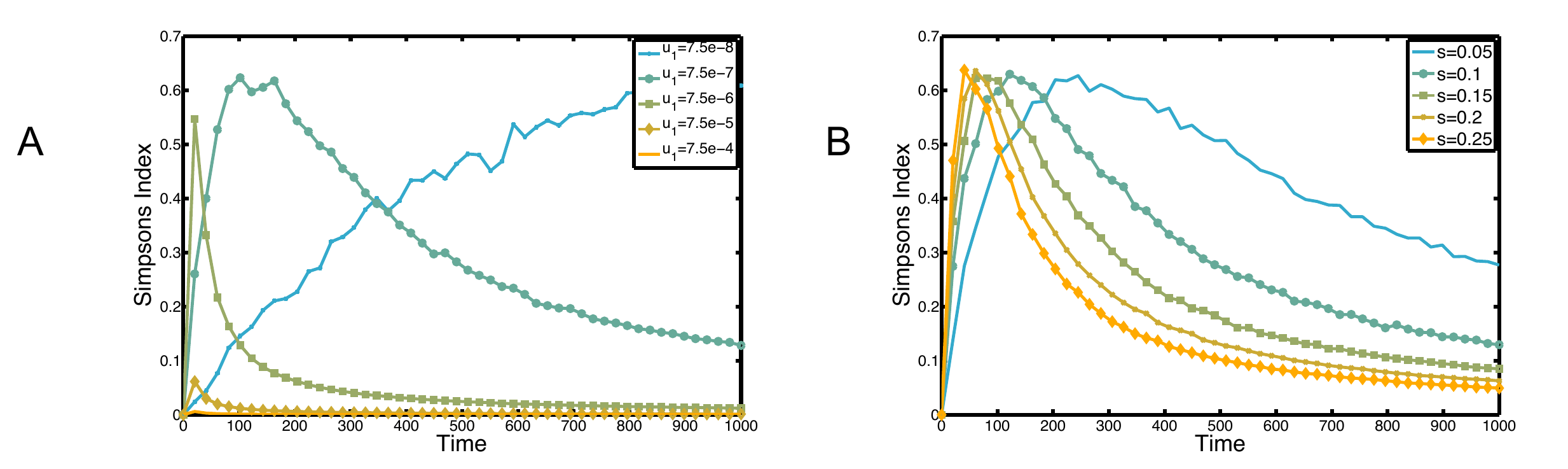} 
   \caption{{\bf Time-dependence of non-spatial Simpson's Index $R(t)$.} The temporal evolution of the expected value of the non-spatial Simpson's Index is shown {\bf (A)} for varying values of the mutation rate $u_1$, and {\bf (B)} for varying values of the fitness advantage $s$ of preneoplastic cells over normal cells. In both simulations: $M_1 = M_2 = 500, N = 10^4$, and  $u_1 = 7.5\times 10^{-7}$, $ s = 0.1$ unless specified.}
   \label{fig:NS_Simpson}
\end{figure}
Next, we use Monte Carlo simulations to evaluate \eqref{eq:Simpson2} and study the temporal evolution of Simpson's Index (see Appendix \ref{app:MC} for details on evaluating \eqref{eq:Simpson2}). In Figure \ref{fig:NS_Simpson}, we observe that the index first increases until it reaches a maximum, and then starts decaying in a monotone fashion. Essentially, this result stems from the fact that in the early phase of the model the first few clones are developing and expanding, so samples are becoming more likely to be from the same clone. As more mutations are produced the population diversifies.  Then, Simpson's Index decreases as it becomes less likely for two cells to be from the same family. Note that this is consistent with the result in Proposition \ref{prop_Simpson_large_n}. Figure \ref{fig:NS_Simpson}A illustrates that as the mutation rate $u_1$ increases, this process of establishing mutant families and diversification occurs faster. In particular, the maximum Simpson's Index decreases with increasing mutation rate because the  time periods in which only a single clone exists become shorter. Finally, Figure \ref{fig:NS_Simpson}B shows that as the selective advantage $s$ is increased, the growth of mutant families and diversification occurs faster due to the faster spread of mutant cells. A possible implication of this result is that more aggressive tumors will have a higher level of heterogeneity. 
\section{Spatial measures of heterogeneity}
\label{sec:SpatialHet}

In the clinical setting, the spatial heterogeneity of premalignant tissues poses considerable challenges.  It is standard practice to take one biopsy sample from an arbitrary location in a suspected premalignant tissue, and to use molecular information from this sample to determine whether it is (pre)cancerous as well as its specific cancer sub-type, if applicable.  This information is used to 
help guide the diagnosis, prediction of prognosis and treatment strategies.   Due to the heterogeneity of premalignant tissue, such a single-biopsy approach may lead to incorrect subtype labeling or diagnoses, and subsequently to  suboptimal therapeutic measures. For example, the spatial extent of this clone is unknown and thus surgical excision or prognosis prediction may be difficult. In view of these issue, the analysis of several biopsies across the tumor mass upon excision seems necessary. However, this raises another question of the {\it length-scale of heterogeneity}: how fine or coarse should the spatial sampling be, i.e.\ how many sections are required for a representative genetic fingerprint of the heterogeneous tissue?\


In order to gain insight into these issues, we focus here on two specific clinical questions and introduce corresponding measures of spatial heterogeneity. 
\begin{itemize}
\item Question 1: Given a region of premalignant tissue, what is the expected length-scale of heterogeneity? (i.e. how far apart should biopsy samples be taken?)  
\item Question 2: Provided that only a single point biopsy is available, what is the expected size of the clone present at the biopsy?  
\end{itemize}
Before we introduce analytical expressions for these two measures of spatial heterogeneity $I_1$ and $I_2$,  we introduce notation that will be useful below.  Suppose two type-1  mutations occur at space-time points $(x_0,t_1)$ and $(y_0,t_2)$, respectively. Then the  two clones will collide at time
$$
t_*=\frac{t_1+t_2}{2}+\frac{||x_0-y_0||}{2c_d}.
$$
Define the vector $v=(y-x)/||y-x||$.  Then the first interaction between the two clones occurs at location
$$
v_*=x_0+c_dv(t_*-t_1)=y_0-c_dv(t_*-t_2).
$$
Next define the half-spaces
$$
H_+=\{x\in\mathbb{R}^d:\langle x,v_*\rangle>0\}\;\mbox{ and }\;H_-=\{x\in\mathbb{R}^d:\langle x,v_*\rangle<0\},
$$ where $\langle \cdot, \cdot \rangle$ is the inner product in $\R^d$. If $x_0\in H_+$ and $y_0\in H_-$ then the region of space influenced by the mutation that occurred at $(x_0,t_1)$ is
$$
B_{x_0,c_d(t-t_1)}\cap H_+,
$$
and similarly the region influenced by the mutation that occurred at $(y_0,t_2)$ is given by
$$
B_{y_0,c_d(t-t_2)}\cap H_-.
$$
Note that we still have
$$
\chi_t=B_{x_0,c_d(t-t_1)}\cup B_{y_0,c_d(t-t_2)}
$$
but we have decomposed $\chi_t$ into regions influenced by the two distinct mutations.

\subsection{Spatial measure $I_1$: length scale of heterogeneity} 
\label{sec:I1}
A mutation at point $(t_i, x_i)$ generates a ball $B_{x_i, c_d(s) (t-t_i)}$ growing linearly in $t$. Thus barring interference, at time $t>t_i$, the type 1 family is of size 
\begin{align}
\label{eq:family_size}
Y_{1,i}(t)={ \gamma_d c_d^d(s)} \ob{t-t_i}^{d},
\end{align}
where $\gamma_d$ is the volume of the $d$-dimensional unit sphere. 
%
To determine the length-scale of spatial heterogeneity,  consider a fixed distance $r>0$ and pick two cells separated by $r$ uniformly at random.  We define $I_1(r,t)$ to be the probability that these two cells are genetically identical (from the same mutant clonal expansion) at time $t$. The functional dependence of $I_1(r,t)$ on $r$  provides an estimate of the length scale of heterogeneity and thus may provide guidance on sampling procedures.  For example, a suggested sampling distance $r_{50} \equiv \{\mbox{argmin}_{r>0} I_1(r,t) < 0.5 \} $  between biopsies would ensure that sampled clones would be genetically different from neighboring samples 50 percent of the time.  The measure $I_1(r,t)$ is a spatial analog of the Simpson's Index.

The actual analysis of $I_1(r,t)$ is quite technical so we will leave the details to the Appendix for interested readers.  However, here we will provide some intuition for our approach, and also provide some graphs demonstrating the dependence of $I_1$ on parameters and time.  We will also provide some comparisons between our analysis (based on the mesoscopic model approximation) and simulations of $I_1$ in the full cell-based stochastic evolutionary model.

{\it Idea behind calculation.} Let $a,b$ be the positions of two cell samples taken at time $t_0$ and assume that $\|a-b\|=r$. 
Define $D_{ab}$ as the event that the cells at positions $a$ and $b$ are genetically different at time $t$. Calculations in sections \ref{sec:I1_1D} and \ref{sec:I1_2D} of the Appendix demonstrate that $\P(D_{ab})$ only depends on the distance between the samples $\|a-b\|$, as long as $c_dt+r\ll L$.  Thus, conditioning on the location of $a$ and $b$ we can conclude that $I_1(r,t) = 1 - \P(D_{ab})$.

Next we discuss the idea behind calculating $\P(D_{ab})$.  Recall that if two clones meet, then each continues to spread in all directions away from the interacting clone. We will denote a cell in position $x$ at time $t$ by the coordinate $(x,t)$. $V_a(t_0)$ and $V_b(t_0)$ are the space-time regions in which a mutation can influence the genetic state of the samples at $a$ and $b$, respectively. The union of these regions $(V_a(t_0) \cup V_b(t_0))$ represents the space-time region in which mutations can influence the genetic state of the samples examined at locations $a$ or $b$ at time $t_0$.
Let $E(A)$ be the number of mutations that occur in a region $A$ and $E_k$  the event $\{E(V_a(t_0) \cup V_b(t_0))=k$\}.  Then, the event $D_{ab}$ can be divided into sub-events according to how many mutations have occurred in the spacetime region $V_a(t_0) \cup V_b(t_0)$
$$
\P(D_{ab}) = \sum_{i=1}^{\infty} \P(D_{ab} \cap E_i)
$$


 The following simple calculation demonstrates that the probability of more than two type-1 mutations occurring in the region $V_a(t_0) \cup V_b(t_0) $ is small for the carcinogenesis setting.  First, we note that the volume $ | V_a(t_0) \cup V_b(t_0)| $ is bounded above by $| V_a(t_0)|+|V_b(t_0) |$.  In 1D this sum of volumes is $c_1(s) t_0^2$ and in 2D it is $\frac{2 \pi}{3} c_2(s)^2 t_o^3$, where $c_1(s)$ and $c_2(s)$ are the spreading speeds of the single mutant clones in dimensions $1$ and $2$ respectively, provided in Section \ref{sec:model}.  Since type-1 mutations arrive into $V_a(t_0) \cup V_b(t_0)$ as a Poisson process with rate $u_1 s$, the number of mutations in $V_a(t_0)\cup V_b(t_0)$ is stochastically dominated by a Poisson random variable with rate $\lambda = u_1 s^2 t_0^2 $ in 1D and 
 $$
 \lambda = u_1 s t_0^3 \frac{4 \pi s}{\log(1/s)} \frac{2 \pi}{3}
 $$ 
 in 2D.  We note that $t_0$ is a time prior to carcinogenesis when premalignant tissue is sampled, and in our model tumor initiation occurs at time $\sigma_2$ when the first successful cell with two mutations arises.   In \cite{DuFoLe15} we found that the appropriate time scale of this process is $1/Nu_1 s$.  Replacing $t_0$ by this in the Poisson rate $\lambda$ in each dimension we obtain $\lambda \equiv 2/N^2 u_1$ in 1D and $\lambda \equiv \frac{8 \pi^2}{3 N^3 u_1^2 s \log(1/s)}$ in 2D.
  
We assume the point mutation rate in healthy tissue is within the range of $10^{-7}$ to $10^{-10}$ per base pair per cell division \cite{Nachmann2000, Kumar2002, Baer2007}.  Selection advantages are more difficult to ascertain experimentally but one study has estimated the average advantage $s$ to be approximately 0.001 \cite{Bozic}.  Lastly, the cell population sizes of interest in tissues at risk of initiating cancer are in the range of $10^6$ and upward.  Using these estimates we can easily calculate that the probability of more than two mutations arriving within the region of interest is negligible across all reasonable parameter ranges.  Thus we can approximate $\P(D_{ab})$ with the first two terms of the sum above:
\begin{align}
\label{eq:Dab}
\P(D_{ab}) \approx \P(D_{ab}\cap E_1)+\P(D_{ab}\cap E_2).
\end{align}
The exact calculations for these quantities $\P(D_{ab}\cap E_1)$ and $\P(D_{ab}\cap E_2)$ are provided in the Section \ref{sec:I1_1D} and \ref{sec:I1_2D} of the Appendix.  

{\it Agreement with microscopic model simulations.}  To verify our results we simulated the full cell-based stochastic evolutionary model and compared the spatial measure $I_1(r,t)$ with our derivations from the previous section.   Table \ref{tab:I1} shows the results of these comparisons.  Since the cell-based model is very computationally intensive, only 100 simulations were performed in each set of parameter values; however the Wald confidence intervals are provided for each set of simulations.  In this table we see a close agreement between our theoretical values for $I_1$ based on the mesoscopic model and the simulations of $I_1$ based on the microscopic model. For all cases, $N=10000$ and $r = 10$.  
\begin{table}[h!]
\begin{center}
\caption{Comparison of $I_1(r,t)$ between theory and simulation of the cell-based stochastic model.}
\label{tab:I1}
\begin{tabular}{c|c|c|c|c|c}
$u_1$ & $s$ & $t$ & Theory $I_1(r,t)$ & Simulation $I_1(r,t)$ & 95$\%$ CI \\
\hline
0.001 & 0.01 & 30 & 0.98 & 0.97 & (0.94, 1)\\
0.001 & 0.01 & 40 & 0.96 & 0.92 & (0.87, 0.97)\\
0.001 & 0.01 & 50 & 0.94 & 0.9 & (0.84, 0.96)\\
0.001 & 0.01 & 60 & 0.90 & 0.9 & (0.84, 0.96)\\
0.001 & 0.01 & 70 & 0.88 & 0.83 & (0.76, 0.90)\\
0.0001 & 0.1 & 30 & 0.9 & 0.95 & (0.91, 0.99) \\
0.0001 & 0.1 & 40 & 0.84 & 0.91 & (0.85, 0.97) \\
0.0001 & 0.1 & 50 & 0.8 & 0.84 & (0.77, 0.91) \\
0.0001 & 0.1 & 60 & 0.79 & 0.8 & (0.72, 0.88) \\

\end{tabular}
\end{center}
\end{table}
\vspace{0.1in}

\begin{figure}[htbp]
\begin{center}
\includegraphics[width=6in]{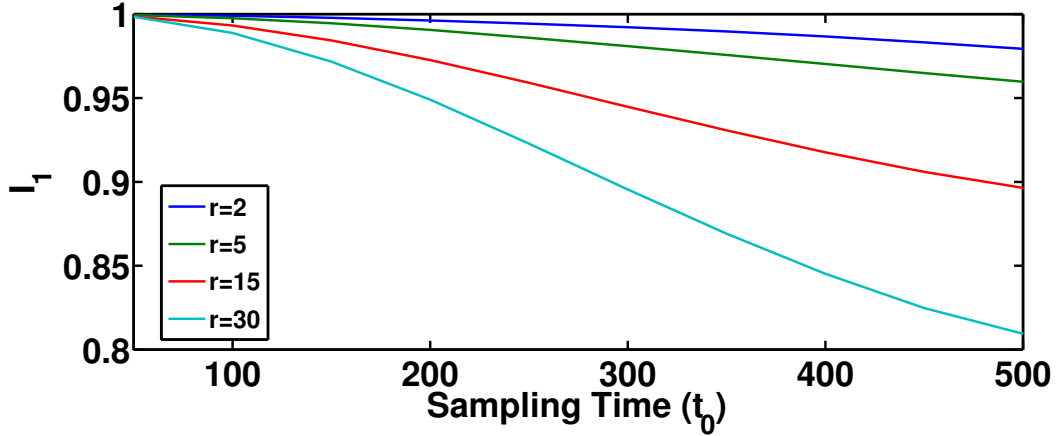}
\caption{$I_1$ in 2D as a function of sampling time $t_0$. We vary the sampling radius $r$ and set $s=0.01$ and $u_1=1e-5$, so the mutation rate is $1e-7$. We also set the mutant growth rate $c_d=0.25$. }
\label{figure:varyt0}
\end{center}
\end{figure}

\begin{figure}[htbp]
\begin{center}
\includegraphics[width=6in]{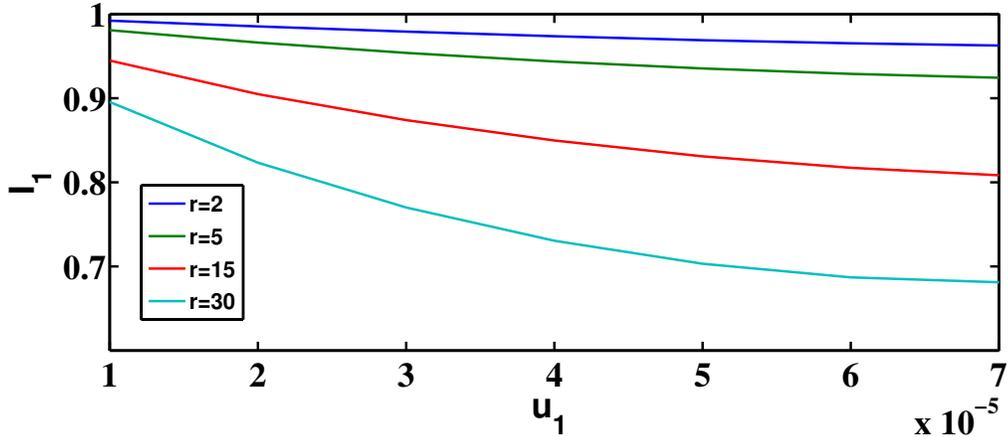}
\caption{$I_1$ in 2D as a function of $u_1$, which contributes to the mutation rate. Mutations arise according to a Poisson process with rate $u_1s$, and we set $s=0.01$  We vary the sampling radius $r$ and set the sampling time $t_0=300$ and the mutant growth rate $c_d=0.25$. }
\label{figure:varyu1}
\end{center}
\end{figure}

In Figures \ref{figure:varyt0} and \ref{figure:varyu1} we demonstrate how the spatial measure of heterogeneity varies in time for different parameters.  In Figure $\ref{figure:varyt0}$ we observe that as time increases $I_1(r,t)$ decreases; this reflects the clonal expansion of existing mutant families leading to an increase in heterogeneity over time.  In addition, as the distance $r$ increases the probability of the two samples being genetically the same decreases, as expected.  Figure $\ref{figure:varyu1}$ shows this result as a function of $u_1$.  As the mutation rate increases, $I_1(r,t)$ decreases since the heterogeneity of the tissue increases with more mutant clones emerging.  
In both figures we see that sensitivity of $I_1$ to parameter changes increases as $r$ increases. This is natural since the likelihood that two cells are identical is more likely to be altered the further apart the two cells are.

\newpage
\subsection{Spatial measure $I_2$: extent of a premalignant lesion}
\label{sec:I2}

Next, suppose we have obtained a premalignant (type-1) biopsy at a single point in the tissue at time $t$.  We would like to estimate the expected size of the corresponding clone. In particular we define $I_2(r,t)$ to be the  probability that an arbitrarily sampled cell at distance $r$ is from the same clone as the original sample.

In order to study $I_2$ it is necessary to define the concept of `size-biased pick' from a sequence of random variables $\{X_i\}_{i\geq 1}$.
\begin{definition}\label{sbp}
A {\it size-biased pick} from the sequence $\ob{X_i}$ is a random variable $X_{[1]}$ such that $$\cP{X_{[1]}=X_i}{X_1, \ldots X_n}= \frac{X_i}{X_1+\ldots +X_n}.$$ 
\end{definition}

Suppose that by time $t$ there have been successful mutations at space time points $\{(x_i,t_i)\}_{i=1}^N$ initiating populations $C_{1,i}$, $1\leq i\leq N$. 
In the model description we stated that the assumptions (A0)-(A3) hold throughout the paper. A consequence of these assumptions, proved in Theorem 4 of \cite{DuFoLe15}, is that overlaps between distinct type-1 cell is unlikely.  Thus we assume here that $1\leq i\leq N$, $C_{1,i}=B_{x_i,c_d(t-t_i)}.$

In order to calculate $I_2$, we first choose a clone $C_{[1]}$ via size biased pick from the different clones $\{C_{1,i}\}$.  The radius of this pick is denoted $R_{[1]}$.    For ease of notation we will make the following substitution throughout the rest of the section $C=C_{[1]}$ and $R=R_{[1]}$.  We next choose a point, $p_1$,  at random from the picked clone $C$.  We choose a  second point $p_2$, at random a distance $r$ away from $p_1$.  In other words, the point $p_2$ is chosen at random from the circle
$$
\mathcal{S}=\{x\in\mathbb{R}^2:|x-p_1|=r\}.
$$
  To calculate $I_2$ we are interested in determining the probability that $p_2$ is contained in $C$. More specifically,   let us denote the center of $C$ by $x_o$.   It is useful to define $X=|p_1-x_o|$ which is a random variable with state space $[0,R]$.  The heterogeneity measure $I_2(r,t)$ is given by:
  \begin{align}
  \label{I2}
I_2(r,t) \equiv P(p_2\in C)=E\left[P\left(p_2 \in C | R,X\right)\right].
\end{align}

  The following two properties are useful in determining $I_2$:
\begin{enumerate}[(i)]
\item If $X+r\leq R$ then $\mathcal{S}\subset C$. To see this consider $z\in \mathcal{S}$ then
\begin{align*}
|z-x_o|&= |z-p_1+p_1-x_o|\leq |z-p_1|+|p_1-x_o|\leq r+X\leq R.
\end{align*}
\item If $R+X<r$ then $\mathcal{S}\cap C = \emptyset$. To see this take $z\in C$ which of course implies $|z-x_o|\leq R$ and thus
$$
|z-p_1| = |z-x_o+x_o-p_1|\leq R+ X<r.
$$
\end{enumerate}
We then have that 
\begin{align}
\label{in1}
P(p_2\in C | R,X) = \begin{cases} 0,&\enskip R+X<r\\
1,&\enskip X+r\leq R\\
\phi(X,R),&\enskip\mbox{ otherwise.}\end{cases}
\end{align}
We can use the cosine rule to see that 
\begin{equation}
\label{eq:phi_I2}
\phi(X,R)=\frac{1}{\pi}\cos^{-1}\left(\frac{X^2+r^2-R^2}{2Xr}\right).
\end{equation}
Substituting expressions \eqref{in1} and \eqref{eq:phi_I2} into \eqref{I2} results in a formula for $I_2$ that can be easily approximated via Monte Carlo simulation.  Details of this procedure are provided in Appendix \ref{sec:I2calc}.\\

The heterogeneity measure $I_2$ is designed to be an estimate of the extent of a premalignant lesion that has already been detected via one point biopsy.  Thus it is of interest to determine the value of $I_2$ at the time of detection of the premalignant condition (which itself may be random).  We hypothesize that detection of the premalignancy may occur at a random time $\tau$ which occurs with a rate proportional to the total man-hours of premalignant lesions.  In other words, detection of the condition is driven by the size and duration of premalignant lesion presence.  Let us define $\tau$ with the following:
\begin{align}
\label{eq:tau_def}
\P\left(\tau>t\right)=\E\exp\left(-\mu\int_0^t|\chi_t|dt\right),
\end{align}
where we recall that $|\chi_t|$ is the volume of type-1 cells at time $t$. Display \eqref{eq:tau_def} tell us that
detection occurs at rate $\mu$. Note we assume that (A1-A3) hold with $u_2s$ replaced by $\mu$.

In Section \ref{sec:I2calc} of the Appendix we also develop a numerical approach for estimating $I_2$ at the random detection time $\tau$. Interestingly enough, it is computationally easier to compute $I_2(r,\tau)$ than $I_2(r,t)$.

\begin{figure}[htbp] 
   \centering
   \includegraphics[width=6in]{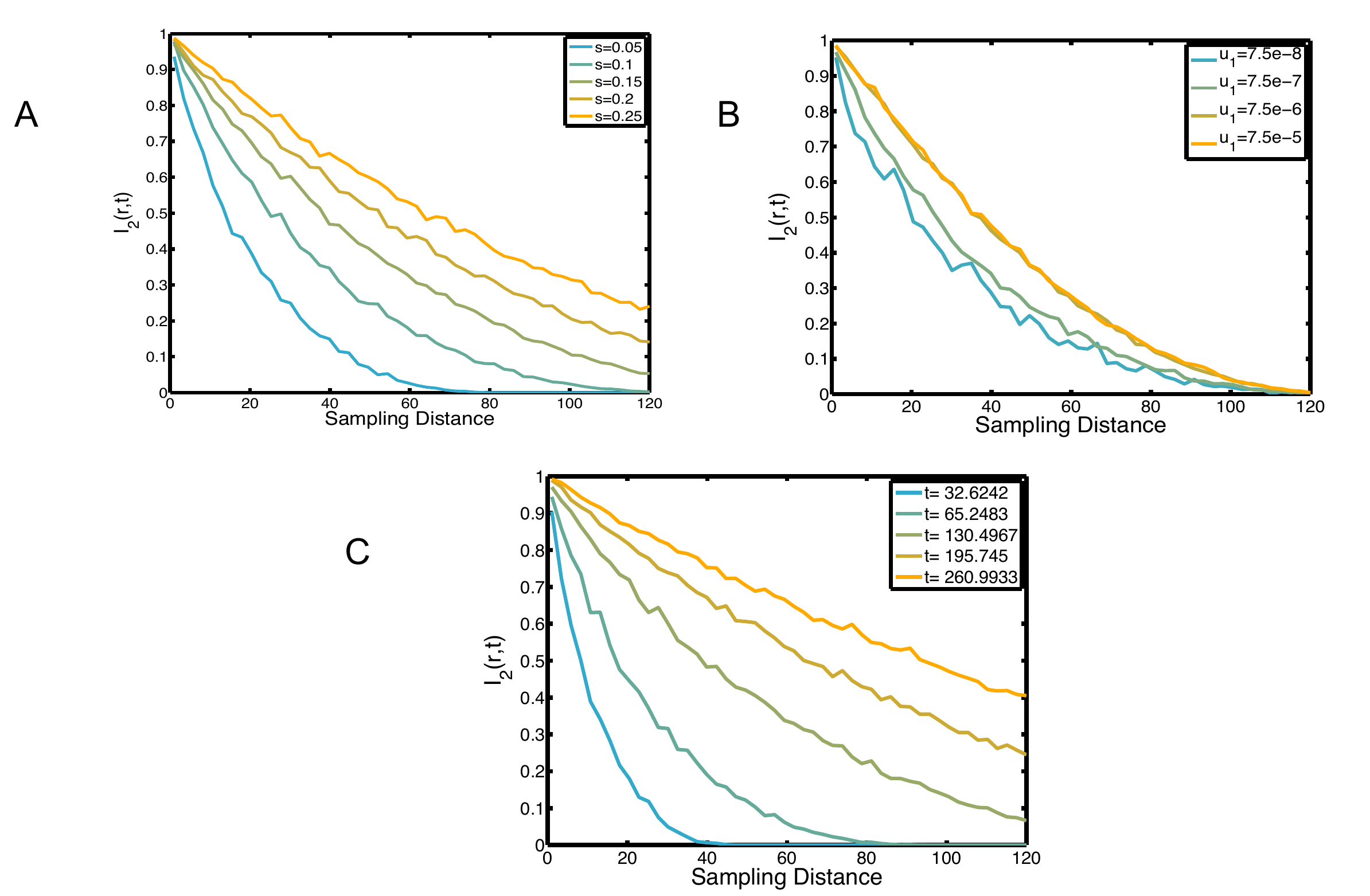} 
   \caption{Plot of $I_2(r,t)$ in 2D as a function of sampling radius for (A) varying selection strength, $s$, (B) varying $u_1$, and (C) varying $t$. In all panels $N=2e5$ and $1e4$ Monte Carlo simulations are performed. Unless varied, $s=0.1$, $u_1=7.5e-7$ and $t$ is the median of the detection time $\tau$ with $\mu=2e-6$.}
   \label{fig:I2t}
\end{figure}

\begin{figure}[htbp] 
   \centering
   \includegraphics[width=6in]{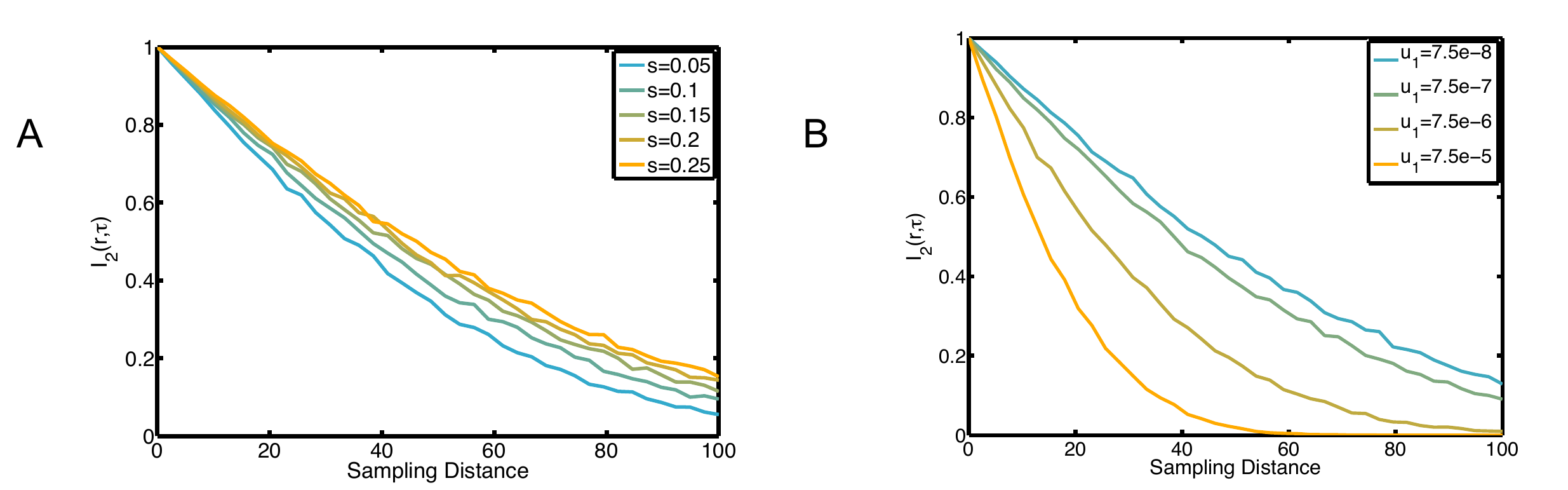} 
   \caption{Plot of $I_2(r,\tau)$ in 2D as a function of sampling radius. In panel (A) we vary the selection strength, $s$ and in panel B we vary $u_1$. In all panels $N=2e5$, we use $1e4$ Monte Carlo simulations and for the random detection time $\tau$ we use $\mu=2e-6$. If not mentioned we set $s=0.1$, $u_1=7.5e-7$.}
   \label{fig:I2tau}
\end{figure}

{\it Numerical examples.} In Figure \ref{fig:I2t}, $I_2(r,t)$ is plotted as a function of $r$ for various values of $u_1$, $s$ and $t$.  Figure \ref{fig:I2tau} shows analogous plots of $I_2(r,\tau)$ at the random detection time.  Comparing Figures \ref{fig:I2t} and \ref{fig:I2tau} we observe an interesting phenomenon. In particular when looking at $I_2$ at a fixed time  in Figure \ref{fig:I2t} we see that for each $r$ and $t$, $I_2$ is an increasing function of both $u_1$ and $s$. This makes sense if we consider the system at a fixed time. Then increasing the mutation rate will increase the expected growing time of any clones, i.e., they are more likely to be born earlier; therefore any clone we select is likely to be larger, so it is more likely that the second point selected a distance $r$ away will be in the original clone. Similarly increasing $s$  increases the expected size of clones present at time $t$, and thus increases $I_2(r,t)$. However, when we look at Figure \ref{fig:I2tau} we see that $I_2(r,\tau)$ is decreasing in $u_1$. Interestingly by observing the process at the random time $\tau$ we flip the dependence on the parameters $u_1$. This phenomenon results from the fact that increasing the mutation rate allows for detection to be caused by multiple clones, which will therefore be smaller at detection than if the detection were driven by a single clone.

\section{Discussion} \label{sec:sum}

In this work we have analyzed and examined several measures of heterogeneity in a spatially structured model of carcinogenesis from healthy tissue.  In particular,  we first derived estimates of the traditionally nonspatial measure of diversity, Simpson's Index, in the premalignant tissue and studied how the Simpson's Index changes in time and varies with parameters.  We observed that as expected, the Simpson's Index decreases over time as more mutants are produced, and that this process occurs faster in a higher mutation rate setting.  The effect of selection overall is also to speed up this process.  We also formulated and analyzed two spatially-dependent measures of population heterogeneity, motivated by clinical questions.  In particular we analyzed a measure ($I_1$) that can identify the length scale of genetic heterogeneity during the carcinogenesis process as well as a measure ($I_2$) that can estimate extent of a surrounding premalignant clone given a premalignant point biopsy.  

We note that in this work we have confined our analysis to a two-step model of carcinogenesis.  The results can be used in a setting where more genetic hits are required for full malignant transformation.  However our heterogeneity estimates would apply to the population of cells with a single mutation and thus can be used in the setting of early stages of carcinogenesis only.  Work on further mutations will be the subject of future work.  

These analyses facilitate a better understanding of how to interpret discrete (in both time and space) samples from a spatially evolving population during carcinogenesis.  For example, the quantity $I_2$ can be calculated to help determine the expected size of a premalignant lesion, given a point biopsy that is premalignant.  In addition the quantity $I_1$ may be used to generate suggestions for optimal sample spacing in situations where multiple biopsies or samples are possible.   Finally, we note that although it is possible to calculate these diversity indices using computational simulation of similar cell-based or agent-based models, it can be extremely computationally onerous to simulate such models for even small sized lattices (100x100 sites) for a lengthy period of time such as during the process of carcinogenesis. Therefore the heterogeneity estimates we derive based on our mesoscopic model in many cases provide the only feasible way to estimate spatial diversity in models living on a larger lattice e.g., 1000x1000 sites or larger. Given the large number of epithelial cells in a small area, realistic simulations to determine statistical properties of diversity measures during carcinogenesis may be completely infeasible. Our results here provide analytical or rapidly computable expressions that enable a detailed assessment of how these heterogeneity measures vary depending on time as well as tissue/genetic parameters such as mutation rate and selective advantage conferred by the genetic alteration.  These tools can be utilized to study how tissue heterogeneity in premalignant conditions varies between sites and tissue types, and thus guide sampling or biopsy procedures across various cancer types.

\appendix

\section{Details for non-spatial Simpson's Index}\label{app:RegSI}

\subsection{Preliminary definitions and results}\label{app:prelim}

To characterize the distributions of the Simpson's Index, we  introduce two definitions.  Let $L_1, L_2, \ldots, L_n$ be independent, identically distributed random variables with distribution $F$. Recall the definition of a sized-biased pick from Definition \ref{sbp} in Section \ref{sec:I2}.
Then we define  a {\it size-biased permutation} as follows. 
\begin{definition}
We call $\ob{L_{[1]}, \ldots L_{[n]}}$ a {\it size-biased permutation} (s.b.p) of the sequence $\ob{L_i}$ if $L_{[1]}$ is a size-biased pick of the sequence, and for $ 2\leq k\leq n$ and 
$$\cP{L_{[k]}=L_j}{L_{[1]},\ldots,L_{[k-1]}; L_1, \ldots L_n}= \frac{L_j \, \mathbbm{1}_{(L_j\neq L_{[i]}, \forall  1\leq i< k)}}{L_1+\ldots+L_n-\ob{L_{[1]}+\ldots+ L_{[k-1]}}}.$$ 
\end{definition}

The following results will be useful.
\begin{proposition}(Proposition 2 in \cite{pitman2012size})
For $1\leq k \leq n$, let $\nu_k$ be the density of $S_k$, the sum of $k$ i.i.d random variables with distribution $F$. Then 
\begin{align}
\P\ob{X_{[1]} \in dx_1, \ldots, X_{[k]}\in dx_k} =& \frac{n!}{(n-k)!}\ob{\prod_{j=1}^k x_j \nu_1(x_j)dx_j }\ldots \nonumber \\ &\ldots \int_0^\infty \nu_{n-k}(s) \prod_{j=1}^k \ob{x_j+\ldots + x_k +s}^{-1} ds. 
\end{align} \qed
\end{proposition}
\begin{corollary}\label{cor1}(Corollary 3 in \cite{pitman2012size})
Let $T_{n-k}= X_{[k+1]}+\ldots+X_{[n]}$ denote the sum of the last $n-k$ terms in an i.i.d s.b.p of length $n$. Then for $k=1,\ldots,n-1$,
\begin{align}
\cP{T_{n-k} \in ds}{T_{n-k+1} = t}= \ob{n-k+1} \frac{t-s}{t} \nu_1(t-s) \frac{\nu_{n-k}(s)}{\nu_{n-k+1}(t)}ds.
\end{align}
\qed
\end{corollary}

\subsection{Conditional expectation}\label{app:condexp}

Recalling that type 1 clones have a linear radial growth rate,  Simpson's index (\ref{Rdef}) can be rewritten explicitly
\begin{align}\label{Rano}
R(t)=\sum_{i=1}^{N_t}\cb{\frac{\ob{1-t_i/t}^{d}}{\sum_{j=1}^{N_t} \ob{1-t_j/t}^{d}}}^2,
\end{align}
where $\set{t_{i}}_{i=1}^{N_t}$ are the points of a Poisson process with constant intensity $Nu_1 st$.  In particular, we note that conditioned on $N_t$, the $t_i$ are i.i.d and 
$$
(t_1/t| N_t) \sim U(0,1).
$$ We define now $X_i:=\ob{1-t_i/t}^{d+1}$ and  let
\begin{align}\label{Tdef}
 T:= \sum_{i=1}^{N_t} X_i,
 \end{align}
which allows us to rewrite (\ref{Rano}) as
\begin{align}\label{Rnew}
R(t)=  \sum_{i=1}^{N_t} \ob{\frac{X_i}{T}}^2.
\end{align}
Note that conditioned on $N_t$, the $X_i$ are i.i.d with $$(X_1| N_t) \sim Beta\ob{\frac{1}{d},1}.$$ To see this, note first that by symmetry $X_i \sim (t_i/t)^{d}$; using characteristic functions, it is then easy to verify that for $X\sim Beta\ob{\alpha,1}$ and $Y\sim U(0,1)$, we have $X\sim Y^n$ if and only if $\alpha= 1/n$.  
Using the above notation and recalling the notion of a size-biased pick in Definition \ref{sbp}, we condition (\ref{Rnew}) on $N_t$ to find
\begin{align}\label{transf}
\cEv{R(t)}{N_t=n}=&\, \sum_{i=1}^{n} {\frac{X_i}{T}} \, \P\ob{X_{[1]}=X_i | X_1, \ldots, X_{n} }\nonumber \\
=&\, \cE{{\frac{X_{[1]}}{T}}}{X_1, \ldots, X_n}.
\end{align}

To compute the conditional expectation of Simpson's Inded $R(t)$, we take the expectation of (\ref{transf}) to find
\begin{align}\label{ER}
\cE{R(t)}{N_t=n} =  \E\ob{\frac{X_{[1]}}{T}}= \int_0^\infty\int_0^\infty \ob{\frac{r}{x}}\,\P\ob{X_{[1]}\in dr, T\in dx }. 
\end{align}
Setting $k=1$, it follows now from Corollary \ref{cor1}
\begin{align}\label{Rcond}
\cE{R(t)}{N_t=n} =& \int_0^\infty\int_0^\infty \ob{\frac{r}{x}}\,\P\ob{T_{n-1}\in d(x-r), T\in dx } \nonumber\\
=& \int_0^\infty\int_0^\infty \ob{\frac{r}{x}} \P(T\in dx) \cP{T_{n-1} \in d(x-r)}{T=x}\nonumber\\
=& n\,\int_0^\infty\int_0^\infty  \ob{\frac{r}{x}}^2 \,  \nu_1(r)\, \nu_{n-1}(x-r)\,dx dr.  
\end{align}
Note that the support of $\nu_1$ is over $[0,1]$, and the support of $\nu_{n-1}$ is over $[0,n]$. 
Now, by definition, $\nu_1$ is the pdf of $Beta\ob{\frac{1}{d},1}$, i.e. $$ \nu_1(x)= \frac{1}{d} x^{\frac{1}{d}-1}.$$ On the other hand, $\nu_{n-1}$ is the density of the sum of $n-1$ i.i.d. $Beta(\frac{1}{d},1)$ random variables, i.e.
$$ \nu_{n-1}(x)= \ob{\nu_1^{\ast(n-1)}}(x).$$

For positive integer $n$ let $S_n=B_1+\ldots + B_n$ where $B_i$ are independent $Beta(\frac{1}{d},1)$ random variables. Finally, from  \eqref{Rcond} we find
\begin{align}
\E[R(t)|N_t=n]=n\E\left[\left(\frac{S_1}{S_n}\right)^2\right],
\end{align}
 $S_n:=B_1+\ldots + B_n$ where $B_i$ are independent $Beta(\frac{1}{d},1)$ random variables.\\

\subsection{Upper bound for variance}\label{app:upboundvar}

We derieve an upper bound for the variance of the conditional Simpson's Index as follows:
\begin{align}
\cb{\cE{ R(t)}{N_t=n}}^2 \leq &  \cE{R^2(t)}{N_t=n} =   \cE{\cb{\cE{\frac{X_{[1]}}{T}}{X_1, \ldots , X_n}}^2}{N_t=n}\nonumber\\
\leq & \cE{\cE{\cb{\frac{X_{[1]}}{T}}^2}{X_1, \ldots, X_n}}{N_t=n}\nonumber \\
= & \cE{\cb{\frac{X_{[1]}}{T}}^2}{N_t=n}\nonumber\\
=& n\int_0^\infty \int_0^\infty  \ob{\frac{r}{x}}^3 \,  \nu_1(r)\, \nu_{n-1}(x-r)\,dx\,dr,
\end{align}
where the second to last equality follows from the fact that the sub-sigma algebra $\sigma(N_t=n)$ is coarser than  $\sigma(X_1, \ldots, X_n)$. 

\subsection{Proof of Proposition \ref{prop_Simpson_large_n}}\label{app:simpnlarge}

\begin{proof}
Let $Y_n=(R(t)|N(t)=n)$, and note that by definition $Y_n\geq 0$. Thus it suffices to show that $\E[Y_n]\to 0$ as $n\to\infty$.  Note that
\begin{align*}
\E[Y_n]&=\frac{1}{n}\E\left[\left(\frac{S_1}{S_n/n}\right)^2\right],
\end{align*}
and by the law of large numbers $S_1/(S_n/n)\to S_1/\E[B_1]$ as $n\to\infty$. Thus if we establish that 
\begin{align}
\label{eq:Simpson_UI}
\sup_{n<\infty}\E\left[\left(\frac{S_1}{S_n/n}\right)^3\right]<\infty,
\end{align}
then by uniform integrability we will have $\E[(S_1/(S_n/n))^2]\to 1$,
and thus $\E[Y_n]\to 0$.

In order to establish \eqref{eq:Simpson_UI} define $S_{2,n}=B_2+\ldots+B_n$ and for $\ve>0$ the event 
$$
A_n=\{S_{2,n}>(1-\ve)(n-1)\E[B_1]\}.
$$ 
We then have that
\begin{align*}
\E\left[\left(\frac{S_1}{S_n/n}\right)^3\right] &=n^3\E\left[\left(\frac{S_1}{S_1+S_{2,n}}\right)^3\right]\\
&=
n^3\E\left[\left(\frac{S_1}{S_1+S_{2,n}}\right)^3;A_n\right]+n^3\E\left[\left(\frac{S_1}{S_1+S_{2,n}}\right)^3;A_n^c\right]\\
&\leq
O(1) + n^3\P(A_n^c).
\end{align*}
From Azuma-Hoeffding inequality we know that there exists a $k$ independent of $n$ such that 
$\P(A_n^c)\leq e^{-kn}$, thus establishing \eqref{eq:Simpson_UI}.
\end{proof}

\subsection{Monte Carlo Simulations}\label{app:MC}
We evaluate the conditional expectation of Simpson's Index for fixed time $t$ using Monte Carlo simulations. Based on the representation in \eqref{eq:Simpson2}, we first generate $M$ independent copies of the vector $(S_1, S_n)$ denoted by $\{(S_1^{(i)},S_n^{(i)})\}_{i=1}^M$, and form the estimator
$$
\hat{\mu}(n,M)=\frac{n}{M}\sum_{i=1}^M\left(\frac{S_1^{(i)}}{S_n^{(i)}}\right)^2,
$$
which satisfies $\E[\hat{\mu}(n,M)]=\E[R(t)|N_t=n]$ and $Var[\hat{\mu}(n,M)]=O(1/M)$. If we simulate $M_1$ copies of $N_t$, denoted by $\{N_t^{(i)}\}_{i=1}^{M_1}$ and then for each realization $N_t=n$ we form the estimator $\hat{\mu}(n,M_2)$ then we have an unbiased  estimator of $\E[R(t)]$ via
$$
\hat{R}(M_1,M_2)=\frac{1}{M_1}\sum_{j=1}^{M_1}\sum_{n=0}^{\infty}1_{\{N_t^{(j)}=n\}}\hat{\mu}(n,M_2),
$$
since $N_t^{(j)}$ is independent of $\hat{\mu}(n,M_2)$.
Note that simulating the mesoscopic model $M$ times and averaging $R(t)$ over those simulations is equivalent to using the estimator $\hat{R}(M,1)$.

\section{$I_1$ Calculations}
Recall from Section \ref{sec:I1} that $I(r,t)$ is approximated by \eqref{eq:Dab}. It is therefore necessary to calculate $\P(D_{ab}\cap E_1)$ and $\P(D_{ab}\cap E_2)$. Also recall that $V_x(t_0)$ is the space-time cone centered at $x$ and has radius $c_dt_0$ at time 0 and radius 0 at time $t_0$. For two points $a$ and $b$ in our spatial domain we will be interested in the sets
\begin{align*}
D(r,t_0)&=V_a(t_0)\Delta V_b(t_0)\\
M(r,t_0)&=V_a(t_0)\cap V_b(t_0),
\end{align*}
where $r=\| a-b\|$.
We suppress the dependence on $a$ and $b$ in $D$ and $M$ to emphasize that the volume of these sets depends only on the distance $\|a-b\|$.
Denote the Lebesgue measure of a set $A\in \mathbb{R}^d\times [0,\infty)$ by $|A|$. In order to calculate $I(r,t)$ it will be necessary to compute $|D(r,t_0)|$ and $M(r,t_0)|$. Note that
\begin{align}
\label{eq:sym_diff}
|D(r,t_0)|=2\left(|V_a(t_0)|-|M(r,t_0)|\right).
\end{align}
In the next two subsections we compute $I_1$ in one and two dimensions. For ease of notation we define $\mu=u_1s$. For real number $a$ define $a^+=\max\{a,0\}$.

\subsection{$I_1$ in 1 dimension}\label{sec:I1_1D}
We will first calculate the volumes $|V_a(t_0)|, |M(r,t_0)|$, and from \eqref{eq:sym_diff} $|D(r,t_0)|$.
In one dimension these calculations are simple: $|V_x(t_0)| = t_0^2c_d$, and $|M(r,t_0)| = \frac{[(2t_0c_d-r)^+]^2}{4c_d}$, so we have that:
\begin{align*}
D(r,t_0) = 2t_0^2c_d-\frac{2[(2t_0c_d-r)^+]^2}{4c_d}.
\end{align*}
Note that if $t_0<r/(2c_d)$ then the only way sites $a$ and $b$ are the same at time $t_0$ is if there are zero mutations in $V_a(t_0)\cup V_b(t_0)$, i.e., 
$$
I(r,t_0)=\exp\left(-\mu |V_a(t_0)\cup V_b(t_0)|\right)=\exp\left(-2\mu t_0^2c_d\right).
$$
Thus assume for the remainder of the subsection that $t_0>r/(2c_d)$, in which case
\begin{align*}
& |V_a(t_0) \cup V_b(t_0)| = 2t_0^2c_d-\frac{(2t_0c_d-r)^2}{4c_d}.
\end{align*}

\vspace{3mm}
And since the mutations arise according to a Poisson process with parameter $\mu$,
$$
\P(E_k) = \displaystyle\frac{(\mu|V_a(t_0) \cup V_b(t_0)|)^ke^{-\mu|V_a(t_0) \cup V_b(t_0)|}}{k!}.
 $$
 
From \eqref{eq:Dab} it thus remains to compute $P(D_{ab}|E_1)$ and $P(D_{ab}|E_2)$. Note that if event $E_1$ occurs then $D_{ab}$ can only occur if the single mutation occurs in the set $D(r,t_0)$ and therefore
\begin{align*}
\P(D_{ab} | E_1) = \frac{|D(r,t_0)|}{|V_a(t_0) \cup V_b(t_0)|} = 1-\frac{(2t_0c_d-r)^2}{8t_0^2c_d^2-(2t_0c_d-r)^2}.
\end{align*}

\begin{figure}[htbp]
\begin{center}
\includegraphics[width=3in]{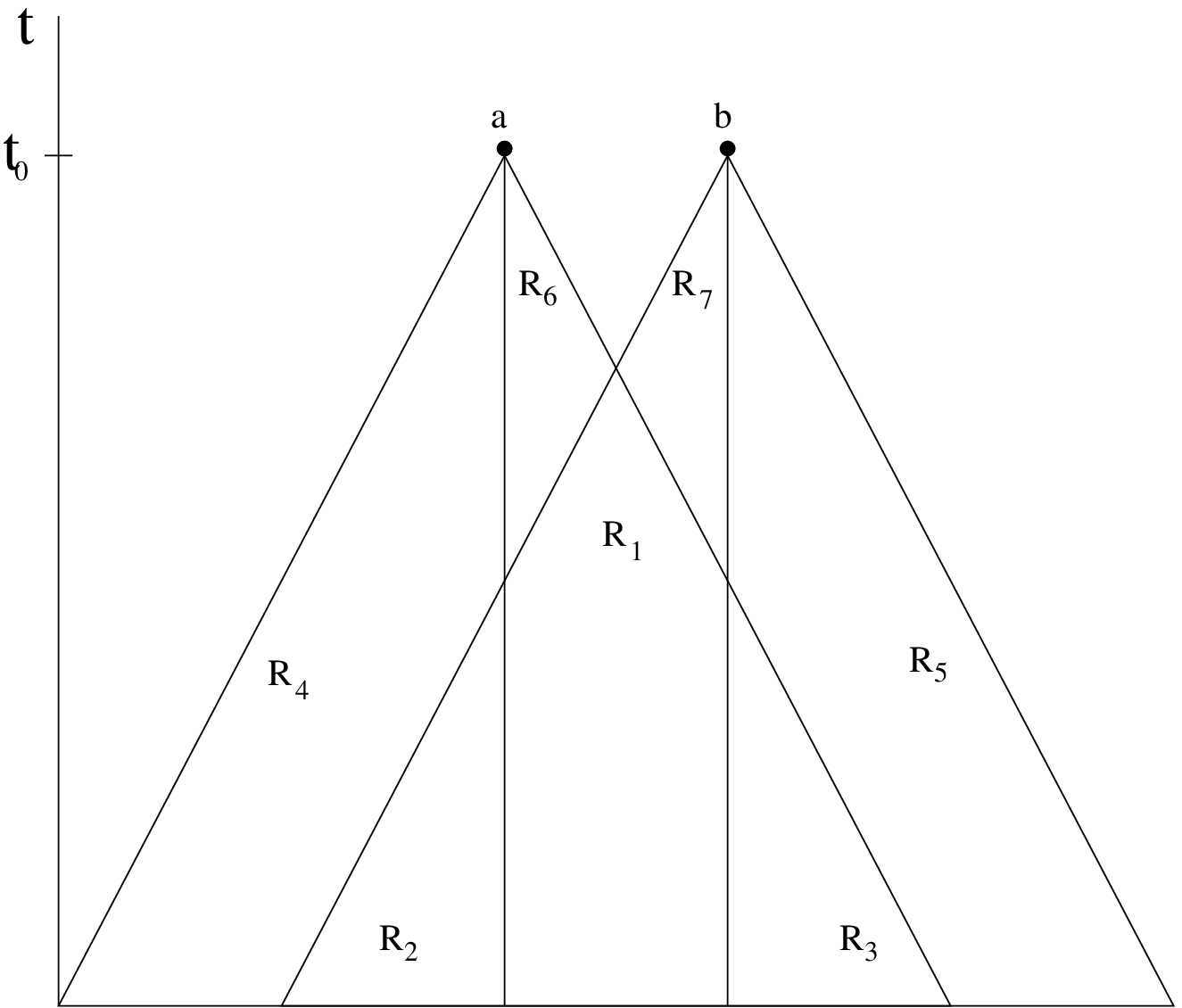}
\caption{Regions}
\label{figure:regions}
\end{center}
\end{figure}

In order to calculate $\P(D_{ab} | E_2)$, we must split $V_a(t_0) \cup V_b(t_0)$ into 7 different regions because the probabilities will differ, depending on where the first mutation occurs (as shown in Figure $\ref{figure:regions}$). By conditioning on $E_2$ we assume that two mutations occur in the space-time region $V_a(t_0)\cup V_b(t_0)$. Denote the space time coordinates of the first mutation by $(x_1,t_1)$.

\vspace{5mm}

If $(x_1,t_1)$ occurs outside of $M(r,t_0)$ but between $a$ and $b$ (i.e. in regions $R_6$ or $R_7$), then the cells will definitely be different, regardless of where the second mutation occurs. However, if the first mutation occurs in $R_i$, $1\leq i \leq 5$, then the location of the second mutation will determine whether the sampled cells are different. Thus each region $R_i$, $1\leq i \leq 5$, will have an associated region $Z_i$ that will be used to calculate $\P(D_{ab} | E_2)$. If the first mutation occurs at the point $(x_1, t_1) \in R_i$, then the shape and size of $Z_i(x_1, t_1)$ depends on $i$ and $(x_1, t_1)$.

\vspace{7mm}

First, we will consider the regions inside $M(r,t_0)$, which are $R_1, R_2,$ and $R_3$. For $i=1,2, 3$, $Z_i(x_1,t_1)$ represents the region in which the occurrence of a second mutation would make the sampled cells different at time $t_0$, i.e. the two clones will meet between $a$ and $b$ and then will each spread to one of the cells. 

\begin{figure}[htbp]
\begin{center}
\begin{subfigure}[b]{0.4\linewidth}

\includegraphics[width=2.5in]{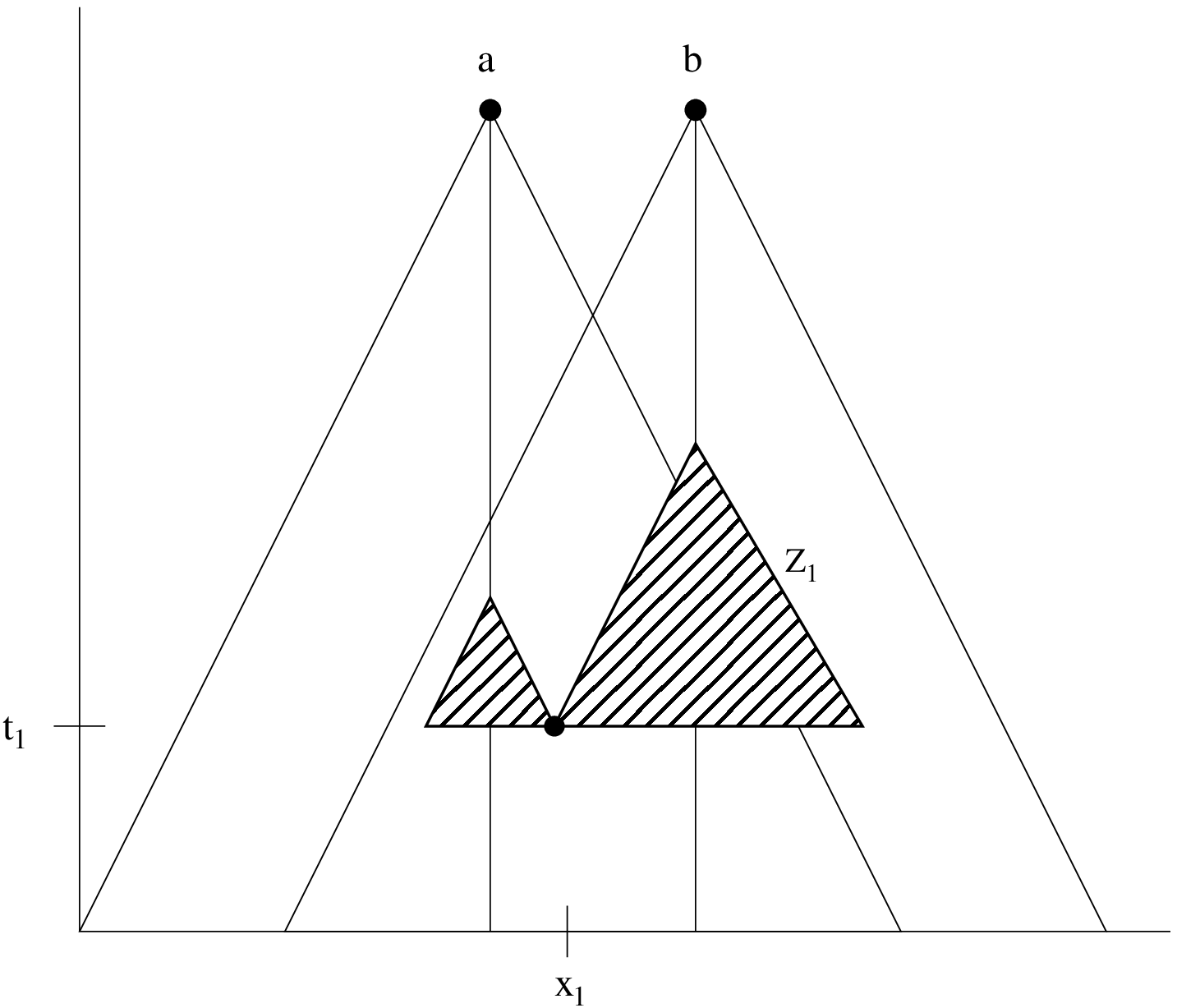}
\caption{$Z_1(x_1,t_1)$}
\label{figure:Z1}
\end{subfigure}
\begin{subfigure}[b]{0.4\linewidth}

\includegraphics[width=2.5in]{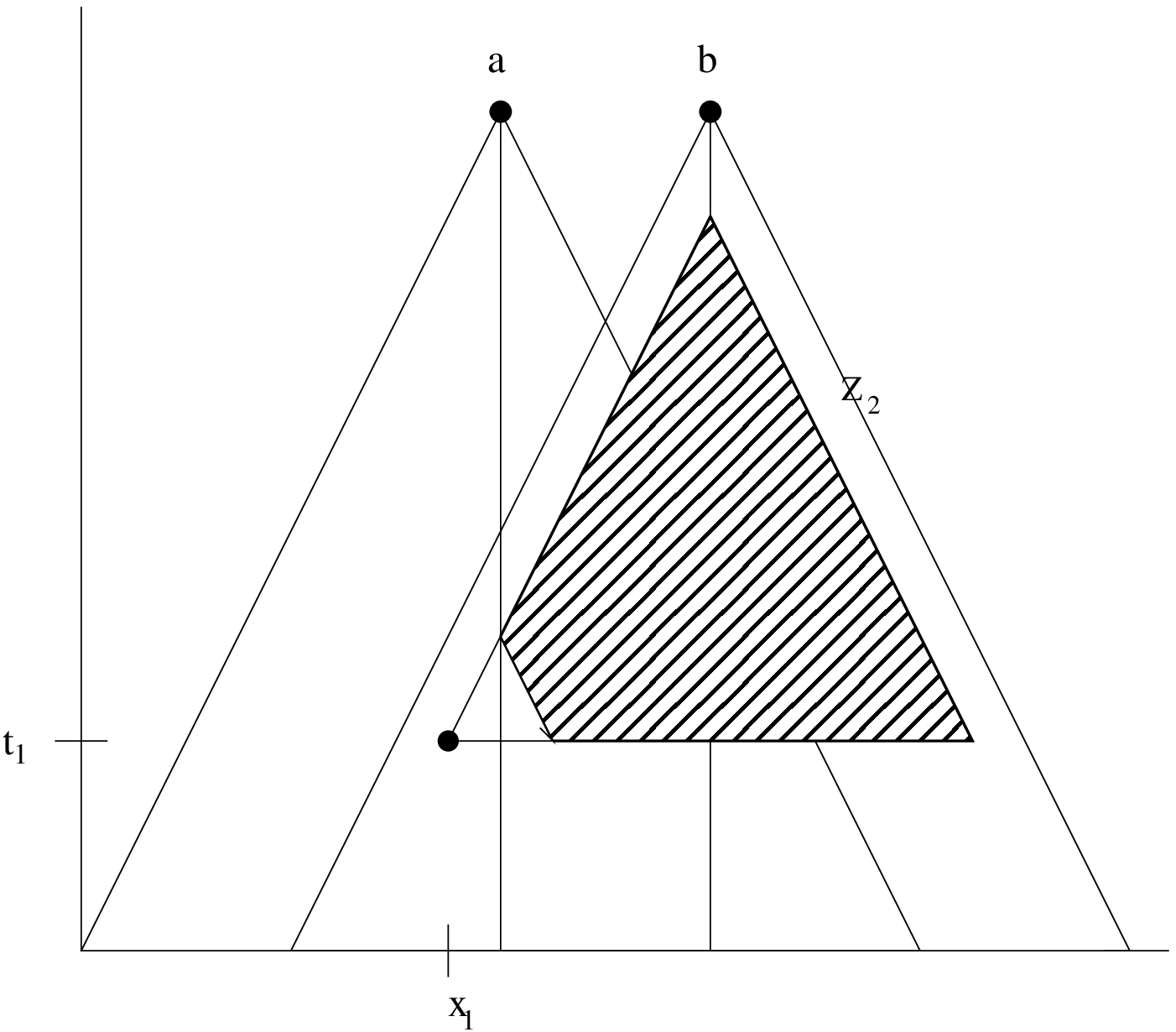}
\caption{$Z_2(x_1,t_1)$}
\label{figure:Z2}
\end{subfigure}

\caption{The region in which the occurrence of a second mutation would make the cells located at $a$ and $b$ different.}

\end{center}
\end{figure}

\begin{figure}[htbp]
\begin{center}
\includegraphics[width=2.5in]{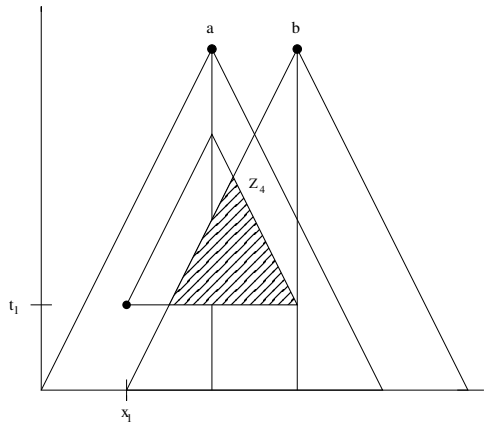}
\caption{$Z_4(x_1,t_1)$: The region in which the occurrence of a second mutation would make the cells located at $a$ and $b$ the same.}
\label{figure:Z4}
\end{center}
\end{figure}

\vspace{5mm}

If $(x_1,t_1) \in R_1$, then $(x_1,t_1)$ is in $M(r,t_0)$ and between $a$ and $b$. In this case, $Z_1(x_1,t_1)$ consists of two triangles, whose upper vertices occur at positions $a$ and $b$ (see Figure $\ref{figure:Z1}$). The base of the triangle on the left is $2(x_1-a)$, and the base of the triangle on the right is $2(b-x_1)$, so the total area of $Z_1(x_1,t_1)$ is $c_d^{-1}[(x_1-a)^2+(b-x_1)^2]$. 

\vspace{5mm}

If $(x_1,t_1) \in R_2$, then $(x_1,t_1)$ is in $M(r,t_0)$ but to the left of $a$. In this case, $Z_2(x_1,t_1)$ is a trapezoidal region. This trapezoidal region can be constructed by taking the triangle whose upper vertex is at position $b$ and subtracting the smaller triangle with upper vertex at position $a$ (see Figure $\ref{figure:Z2}$). The base of the larger triangle is $2(b-x_1)$, and the base of the smaller triangle is $2(a-x_1)$. Hence, the area of $Z_2(x_1,t_1)$ is $c_d^{-1}[(b-x_1)^2-(a-x_1)^2]$. $Z_3(x_1,t_1)$ is constructed analogously to $Z_2(x_1,t_1)$.

\vspace{7mm}

$Z_4(x_1,t_1)$ and $Z_5(x_1,t_1)$ have a slightly different meaning. Given that the first mutation occurs in region 4 or 5, respectively, $Z_4(x_1,t_1)$ and $Z_5(x_1,t_1)$  each represent the region in which the occurrence of a second mutation would make the sampled cells the genetically identical. 

\vspace{5mm}

If $(x_1,t_1) \in R_4$, then $(x_1,t_1)$ is outside of $M(r,t_0)$ and to the left of $a$.  In order for $a$ and $b$ to be the same in this case, the second clone must meet the first clone before it reaches $a$, and the second clone must spread to $b$ before $t_0$. Hence, $Z_4(x_1,t_1)$ is a triangle inside $M(r,t_0)$ (see Figure $\ref{figure:Z4}$). In the next paragraph we will explain how the area of $Z_4(x_1,t_1)$ is calculated.

\vspace{5mm}

The distance between the right vertex of $Z_4(x_1,t_1)$ and $a$ is equal to the distance between $a$ and $x_1$, so the position of that vertex is $a+(a-x_1)=2a-x_1$. Let $V_b'$ be the portion of $V_b$ that falls between the $t$-values $t_1$ and $t_0$. Then we can find the position of the left vertex of $Z_4(x_1,t_1)$ by considering it as the left corner of $V_b'$.  The height of $V_b'$ is $t_0-t_1$, so its base is $2c_d(t_0-t_1)$. Then the left vertex of $V_b'$, and consequently the left vertex of $Z_4(x_1,t_1)$ is $b-c_d(t_0-t_1)$. Hence, the base of $Z_4(x_1,t_1)$ has length $2a-x_1-b+c_d(t_0-t_1)$. Therefore, the area of $Z_4(x_1,t_1)$ is:
$$
\displaystyle\frac{(2a-x_1-b+c_d(t_0-t_1))^2}{4c_d}.
$$  
Analogously, the area of $Z_5(x_1,t_1)$ is: 
$$
\displaystyle\frac{(a-2b+x_1+c_d(t_0-t_1))^2}{4c_d}.
$$

\vspace{5mm}

In summary, we have the following areas:
\begin{align*}
\mbox{If } (x_1,t_1) \in R_1, \mbox{then } |Z_1(x_1,t_1)| = c_d^{-1}[(x_1-a)^2+(b-x_1)^2]\\
\mbox{If } (x_1,t_1) \in R_2, \mbox{then } |Z_2(x_1,t_1)| = c_d^{-1}[(b-x_1)^2-(a-x_1)^2]\\
\mbox{If } (x_1,t_1) \in R_3, \mbox{then } |Z_3(x_1,t_1)| = c_d^{-1}[(x_1-a)^2-(x_1-b)^2]\\ 
\mbox{If } (x_1,t_1) \in R_4, \mbox{then } |Z_4(x_1,t_1)| = \displaystyle\frac{(2a-x_1-b+c_d(t_0-t_1))^2}{4c_d}\\
\mbox{If } (x_1,t_1) \in R_5, \mbox{then } |Z_5(x_1,t_1)| = \displaystyle\frac{(a+c_d(t_0-t_1)-2b+x_1)^2}{4c_d}
\end{align*}

\vspace{5mm}

Let $X_n$ be the position of the $n$th mutation. Then:
\begin{align*}
\P(D_{ab} | E_2) = &\sum\limits_{i=1}^3\P(X_2\in Z_i | X_1 \in R_i)\P(X_1 \in R_i) \\ & + \sum\limits_{i=4}^5\P(X_2\notin Z_i | X_1 \in R_i)\P(X_1 \in R_i)  + \sum\limits_{i=6}^7\P(X_1 \in R_i).
\end{align*}
Thus to calculate $\P(D_{ab})$ it remains to calculate $\P(X_2\in Z_i|X_1\in R_i)$ and $\P(X_1\in R_i)$ for $i\in\{1,\ldots,5\}$.

\begin{figure}[htbp]
\begin{center}
\begin{subfigure}[b]{0.4\linewidth}

\includegraphics[width=2.5in]{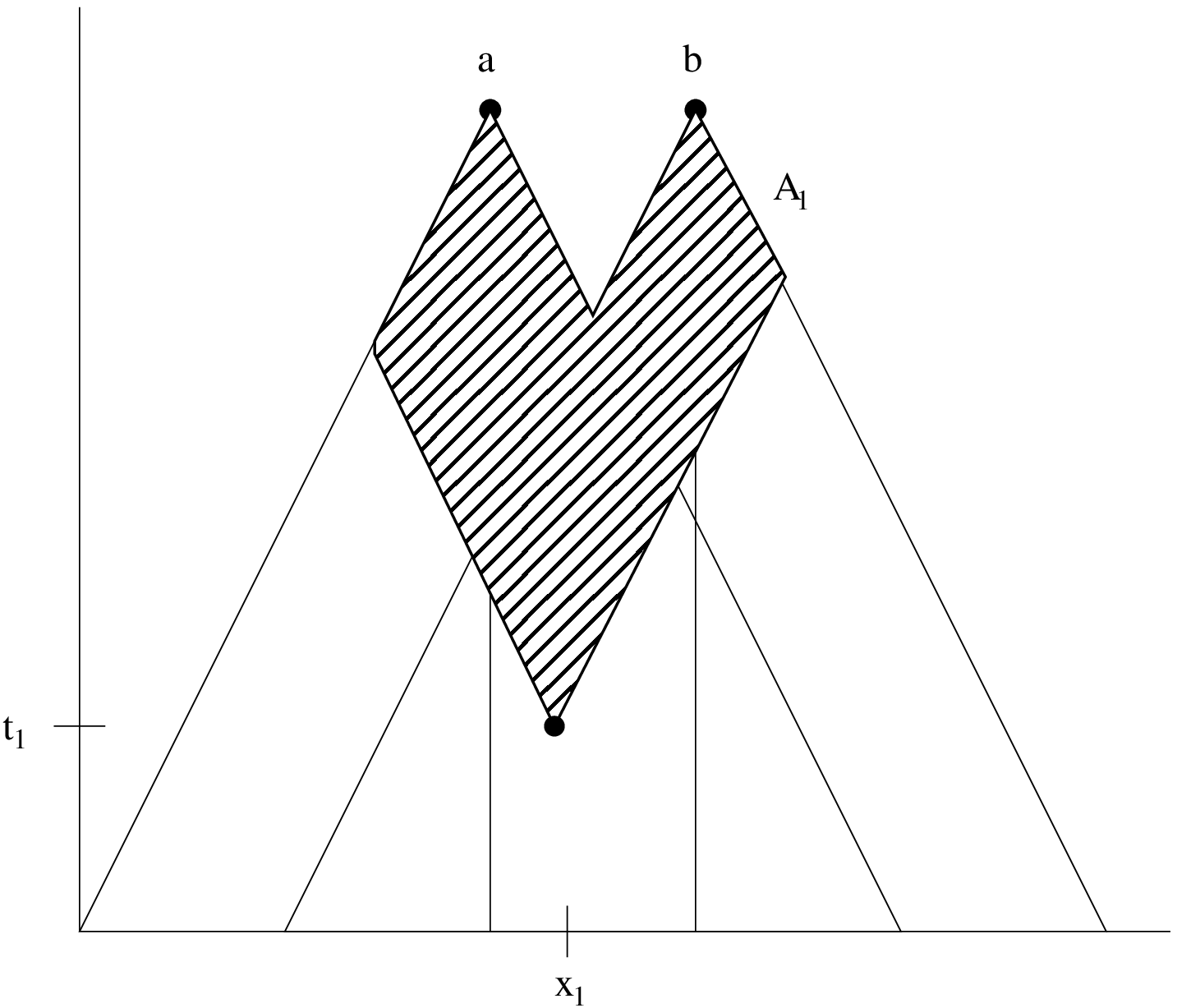}
\caption{$A_1(x_1,t_1)$}
\label{figure:A1}
\end{subfigure}
\begin{subfigure}[b]{0.4\linewidth}

\includegraphics[width=2.5in]{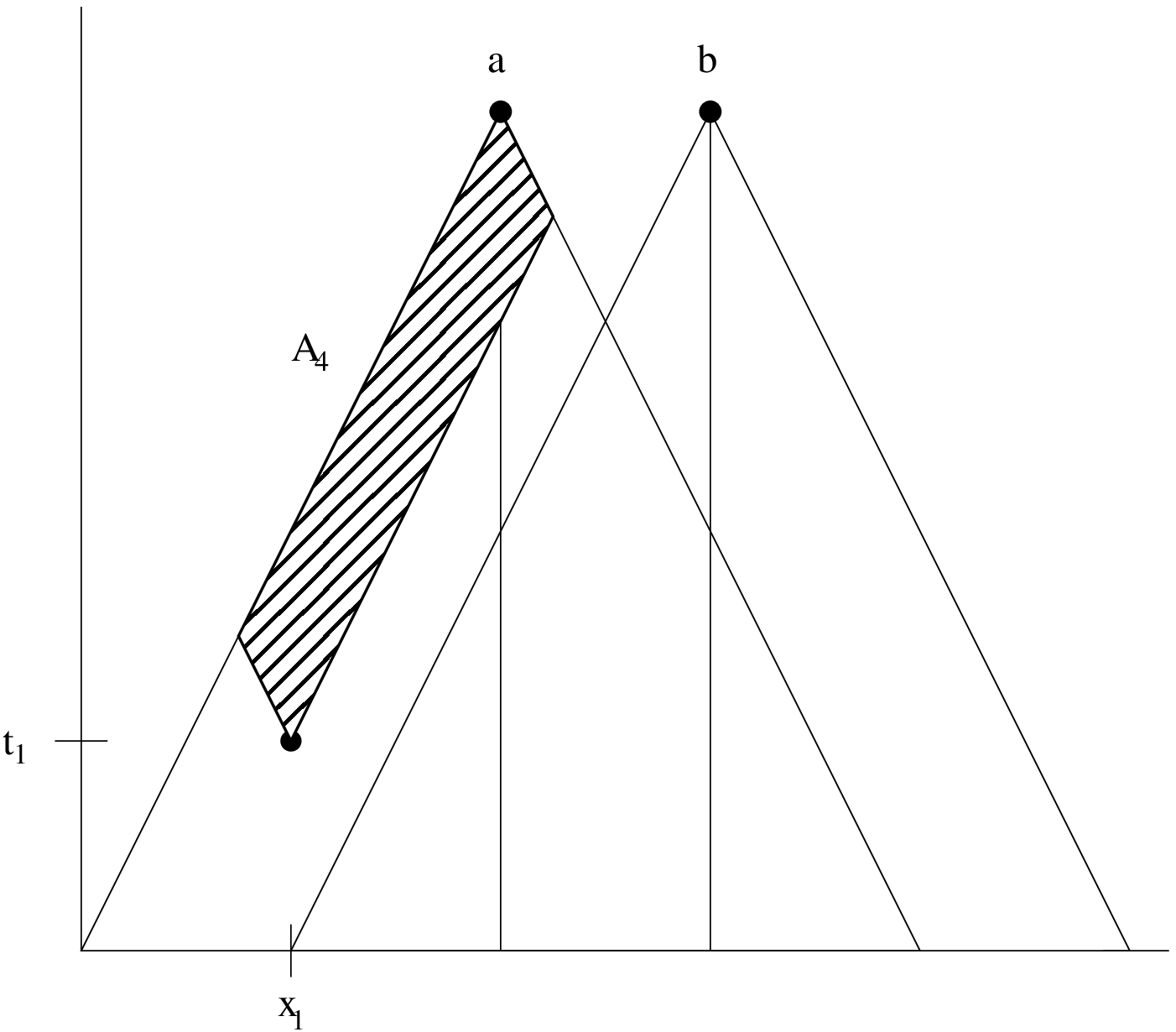}
\caption{$A_4(x_1,t_1)$}
\label{figure:A4}
\end{subfigure}

\caption{The region inside $V_a\cup V_b$ that is affected by a mutation at $(x,t) \in R_i$, and thus is not susceptible to subsequent mutation.}

\end{center}
\end{figure}

Let $A_i(x,t)$ be the region inside $V_a\cup V_b$ that is affected by a mutation at $(x,t) \in R_i$. Since type-1 mutations must occur in cells that have not yet mutated, the second type-1 mutation cannot occur inside $A_i(x,t)$. 

\vspace{3mm}

The area of $A_i(x,t)$ depends on whether (x,t) is in $M(r,t_0), V_a\setminus V_b$, or $V_b\setminus V_a$. The following are the areas $|A_i(x,t)|$, which will be used to calculate $\P(X_2\in Z_i|X_1\in R_i)$:

\begin{align*}
&|A_1(x,t)| = |A_2(x,t)| = |A_3(x,t)| = \\
&c_d(t_0-t+\frac{r}{2c_d})^2- \displaystyle\frac{2r^2+(b-x+c_d(t_0-t))^2+(x-a+c_d(t_0-t))^2}{4c_d}\\
&|A_4(x,t)| = |A_6(x,t)| = c_d(t_0-t)^2 -\displaystyle\frac{(x-a+c_d(t_0-t))^2 + (a-x+c_d(t_0-t))^2}{4c_d}\\
&|A_5(x,t)| = |A_7(x,t)| = c_d(t_0-t)^2 -\displaystyle\frac{(x-b+c_d(t_0-t))^2 + (b-x+c_d(t_0-t))^2}{4c_d}.
\end{align*}

\vspace{3mm}

We will explain how $|A_4(x,t)|$ is calculated and leave out the calculations for $|A_1(x,t)|$ and $|A_5(x,t)|$, which can be done similarly.

$|A_4(x,t)|$ is calculated by taking the area of the truncated triangle $V_a'$ (the portion of $V_a$ that lies between times $t$ and $t_0$) and then subtracting the area of two smaller triangles that are not in $A_4$ (see Figure $\ref{figure:A4}$). The bases of these triangles lie along line $t$, between $x$ and the two lower vertices of $V_a'$. The height of $V_a'$ is $t_0-t$, so the base is $2c_d(t_0-t)$. Hence the lower left vertex of $V_a'$ is at position $a-c_d(t_0-t)$, and the lower right vertex of $V_a'$ is at position $a+c_d(t_0-t)$. Therefore the base of the left small triangle is $x-a+c_d(t_0-t)$, so its area is $\displaystyle\frac{(x-a+c_d(t_0-t))^2}{4c_d}$. The base of the right small triangle is $a+c_d(t_0-t)-x$, so its area is $\displaystyle\frac{(a-x+c_d(t_0-t))^2}{4c_d}$. Since $|V_a'| =  c_d(t_0-t)^2$, we get the area listed above for $A_4$ and $A_6$. 

\vspace{7mm}

If $X_1=(x,t)\in R_i$, then $\P(X_2\in Z_i)=\displaystyle \frac{|Z_i(x,t)|}{|V_a\cup V_b \setminus A_i(x,t)|}$. We can integrate this quantity over the places where the first mutation could have occurred, which is all of $R_i$, and then divide by $|R_i|$ to get: 
\begin{align*}
& \P(X_2\in Z_i | X_1 \in R_i) = \frac{1} {|R_i|}\int_{R_i}\frac{|Z_i(x,t)|}{|V_a\cup V_b \setminus A_i(x,t)|}dxdt.
\end{align*}

\vspace{5mm}

Now it remains to calculate $\P(X_1 \in R_i)$.  Since mutations arrive according to a Poisson process, we have $\P(X_1 \in R_i) = \mu(|R_i|)e^{-\mu(|R_i|)}$, and it suffices to know the following areas:
\begin{align*}
& |R_1| = \displaystyle\frac{(2t_0c_d-r)^2}{4c_d}- \frac{(a-b+c_dt_0)^2}{2c_d}\\
& |R_2| = |R_3| = \displaystyle\frac{(a-b+c_dt_0)^2}{4c_d}\\
& |R_4| = |R_5| = \frac{c_dt_0^2}{2} - \displaystyle\frac{(a-b+c_dt_0)^2}{4c_d}\\
& |R_6| = |R_7| = \frac{r^2}{4c_d}.
\end{align*}

The expression for $|R_2|$ listed above is calculated by considering $R_2$ as a triangle inside $V_b$. The height of $V_b$ is $t_0$, so  the left vertex is at position $b-c_dt_0$. Then the base of $R_2$ is $a-b+c_dt_0$, which means its area is $\displaystyle\frac{(a-b+c_dt_0)^2}{4c_d}$.

\vspace{3mm}

Then we can use $|R_2|$ to calculate $|R_1|$ and $|R_6|$:

 \vspace{3mm}

$|R_1| = |M(r,t_0)| - 2|R_2|$, and $|R_4| = \frac{1}{2}|V_a| - |R_2|$. And the height of $R_6$ is $t_0$ minus the height of $R_2$, so $|R_6|$ simplifies to $\displaystyle\frac{r^2}{4c_d}$. 

\vspace{7mm}

All of the equations above can be used to calculate $\P(D_{ab})$:
\begin{align*}
\P(D_{ab}) \approx & \P(D_{ab} | E_1) \P(E_1) + \P(E_2)\big(\sum\limits_{i=1}^3\P(X_2\in Z_i | X_1 \in R_i)\P(X_1 \in R_i)  \\ & + \sum\limits_{i=4}^5\P(X_2\notin Z_i | X_1 \in R_i)\P(X_1 \in R_i)  + \sum\limits_{i=6}^7\P(X_1 \in R_i)\big).
\end{align*}

\subsection{$I_1$ in 2 dimensions}\label{sec:I1_2D}
Similary to the one dimensional case, we will first calculate $|V_a(t_0)|$ and $|M(r,t_0)|$. In the two dimensional setting this is slightly more difficult. First we know that $|V_a(t_0)|=\pi t_0^3c_d^2/3$, so it remains to find $|M(r,t_0)|$. 
\begin{figure}[htp]
\centering
\scalebox{0.6}{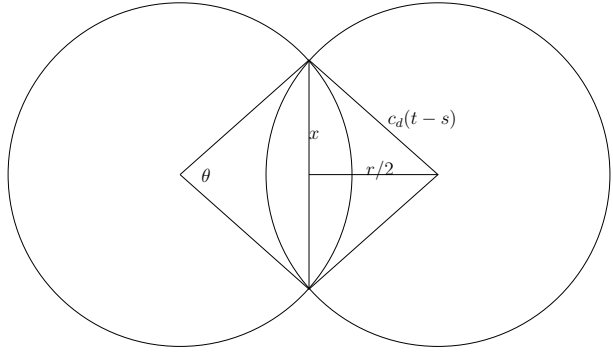}
\caption{Overlap of space time cones at time $s$.}\label{fig:circle_overlap}
\end{figure}

\vspace{3mm}

First observe that if $r>2c_dt_0$ then $M(r,t_0)=\emptyset$, so we only need to calculate $|M(r,t_0)|$ in the case $r<2c_dt_0$.
If we consider the overlap of space-time cones at the fixed time $s\in \left[0,t_0-r/(2c_d)\right]$
then looking at Figure \ref{fig:circle_overlap} it can be seen that half the area of the overlap of their cones at this specific time is given by taking the difference between the area of the circular section with radius $c_d(t-s)$ and angle $\theta$ and twice the area of the triangle with side lengths $x,r/2, c_d(t_0-s)$. The area of the circular section is given by
$$
c_d^2(t_0-s)^2\cos^{-1}\left(\frac{r}{2c_d(t_0-s)}\right),
$$
and twice the area of the triangle is given by
$$
\frac{r}{2}\sqrt{c_d^2(t_0-s)^2-r^2/4}.
$$
Thus the area of overlap between the two cones at time $s$ is given by
$$
a(s)=2\left(c_d^2(t_0-s)^2\cos^{-1}\left(\frac{r}{2c_d(t_0-s)}\right)-\frac{r}{2}\sqrt{c_d^2(t_0-s)^2-r^2/4}\right).
$$
The space-time volume of $M(r,t)$ is therefore given by
\begin{align*}
|M(r,t_0)|=&\int_0^{t_0-r/2c_d}a(s)ds\\
 =&
 2\int_0^{t_0-r/2c_d}\left(c_d^2(t_0-s)^2\cos^{-1}\left(\frac{r}{2c_d(t_0-s)}\right)-\frac{r}{2}\sqrt{c_d^2(t_0-s)^2-r^2/4}\right)ds\\
=&
\frac{2}{c_d}\int_{r/2}^{c_dt_0}y^2\cos^{-1}\left(\frac{r}{2y}\right)dy-\frac{r}{c_d}\int_{r/2}^{c_dt_0}\sqrt{y^2-r^2/4}\,dy.
\end{align*}
Applying integration by parts to the first integral we see that
\begin{align*}
\frac{2}{c_d}\int_{r/2}^{c_dt}y^2\cos^{-1}\left(\frac{r}{2y}\right)dy=
\frac{2c_d^2t_0^3}{3}\cos^{-1}\left(\frac{r}{2c_dt_0}\right)-\frac{r}{3c_d}\int_{r/2}^{c_dt_0}\frac{y^2}{\sqrt{y^2-r^2/4}}dy.
\end{align*}
We thus have that
\begin{align*}
|M(r,t_0)|=&\frac{2c_d^2t_0^3}{3}\cos^{-1}\left(\frac{r}{2c_dt_0}\right)-\frac{r}{6c_d}\int_{r/2}^{c_dt_0}\frac{16y^2-3r^2}{\sqrt{4y^2-r^2}}dy\\
&=
\frac{2c_d^2t_0^3}{3}\cos^{-1}\left(\frac{r}{2c_dt_0}\right)-\frac{rt_0}{3}\sqrt{4c_d^2t_0^2-r^2}-\frac{r^3}{12c_d}\log\left(\frac{r}{2c_dt_0+\sqrt{4c_d^2t_0^2-r^2}}\right),
\end{align*}
which we can combine with \eqref{eq:sym_diff} to see that for $r<2c_dt_0$
\begin{align*}
|D(r,t)|&=\frac{2c_d^2t_0^3}{3}\left(\pi-2\cos^{-1}\left(\frac{r}{2c_dt_0}\right)\right)+\frac{2rt_0}{3}\sqrt{4c_d^2t_0^2-r^2}\\
&\quad+
\frac{r^3}{6c_d}\log\left(\frac{r}{2c_dt_0+\sqrt{4c_d^2t_0^2-r^2}}\right).
\end{align*}

With these calculations we see that
\begin{align*}
|V_a(t_0) \cup V_b(t_0)| &= \frac{2c_d^2t_0^3}{3} \left(\pi - \cos^{-1}\left(\frac{r}{2c_dt_0}\right)\right)+\frac{rt_0}{3}\sqrt{4c_d^2t_0^2-r^2}\\
&\quad+\frac{r^3}{12c_d}\log\left(\frac{r}{2c_dt_0+\sqrt{4c_d^2t_0^2-r^2}}\right),
\end{align*}
We can now explicitly calculate $\P(D_{ab} | E_1) = \frac{|D(r,t_0)|}{|V_a(t_0) \cup V_b(t_0)|}.$ The remainder of this section will deal with the calculation of $\P(D_{ab}|E_2)$.

\vspace{3mm}

The approach here will be slightly different from the one-dimensional case because it is easier to look at the two-dimensional cross sections of $|V_a(t_0) \cup V_b(t_0)|$, rather than the entire three-dimensional space-time cones. Therefore, we will split the cross sections into just two regions, and then when calculating the relevant volumes involved in $I_2$, we will split the regions into multiple cases. In the end, the process is similar, but the setup will be simpler, and then the volume calculations will be more complicated in the two-dimensional setting.

\begin{figure}[htbp]
\begin{center}
\includegraphics[width=3in]{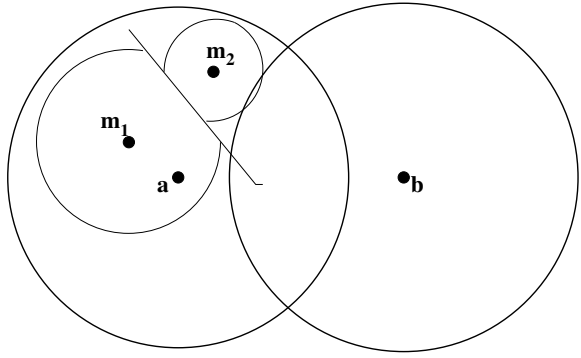}
\caption{When two mutation circles collide, they will continue to expand along the line perpendicular to the line segment joining the two mutation origins.}
\label{figure:interact}
\end{center}
\end{figure}

\vspace{7mm}

If two events occur in $V_a(t_0) \cup V_b(t_0)$, then the probabilities will differ, depending on whether the first event occurs in $M(r,t_0)$ or in $D(r,t_0)$. We will assume that if two mutation circles collide, then we can draw a line through that point, perpendicular to the line segment connecting the two mutations (as show in Figure $\ref{figure:interact}$).  The circles will not extend beyond that line but will continue to expand in all other directions.

\begin{figure}[htbp]
\begin{center}
\includegraphics[width=3in]{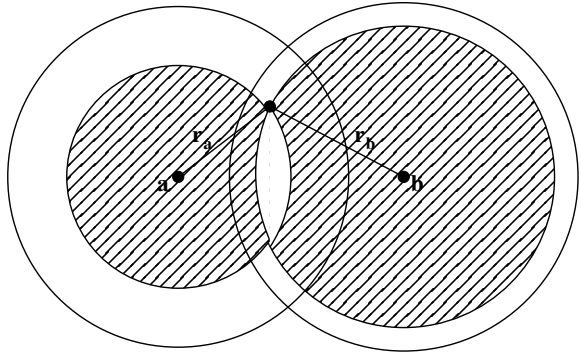}
\caption{The cross-section of the cones $V_a(t_0)$, $V_b(t_0)$, $C_a(t_1)$, and $C_b(t_1)$ at the moment when a mutation occurs in the intersection, $M(r,t_0)$. If a second mutation occurs in the shaded mutation, then the cells located at $a$ and $b$ will be different.}
\label{figure:inters}
\end{center}
\end{figure}

\vspace{7mm}

If the first event occurs in $M(r,t_0)$ at position $(x_1, y_1)$ at time $t_1$, then let $r_a$ be the distance between $(x_1, y_1)$ and $a$, and let $r_b$ be the distance between $(x_1, y_1)$ and $b$. 
Then let $C_a(t_1)$ be the cone centered at $a$ that extends to the edge of the expanding clone, so $C_a(t_1)$ will have radius $r_a$ at time $t_1$ and radius $0$ at time $(t_1 + \frac{r_a}{c_d})$.  Similarly, $C_b(t_1)$ will be the cone centered at $b$ with radius $r_b$ at time $t_1$ and radius $0$ at time $(t_1 + \frac{r_b}{c_d})$. Cross-sections of these cones are shown in Figure \ref{figure:inters}.

\vspace{5mm}

If the second mutation occurs outside of $C_a(t_1) \cup C_b(t_1)$, then the first clone will reach both $a$ and $b$ before interacting with the second clone. If the second mutation occurs in $C_a(t_1) \setminus C_b(t_1)$, then the line dividing the two clones will separate $a$ from $b$, so the second clone will affect $a$, and the first will affect $b$, making the two cells different. Similarly, if the second mutation occurs in $C_b(t_1) \setminus C_a(t_1)$, then the first clone will affect $a$, and the second will affect $b$. 

\vspace{5mm}

However if the second mutation occurs in $C_b(t_1) \cap C_a(t_1)$, then both $a$ and $b$ will be on the same side of the line dividing the mutation circles, so the second clone will affect both $a$ and $b$. Therefore, the two cells will only be different if the second mutation occurs in $C_b(t_1) \triangle C_a(t_1)$.

\begin{figure}[htbp]
\begin{center}
\includegraphics[width=3in]{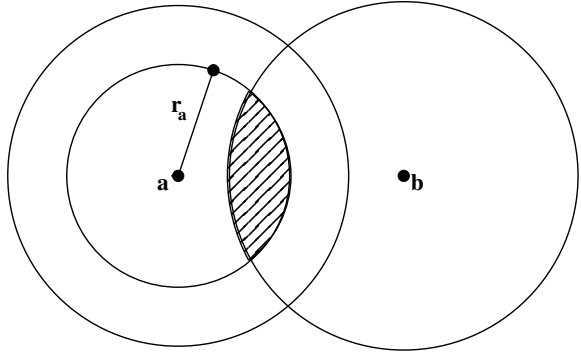}
\caption{The cross-section of the cones $V_a(t_0)$, $V_b(t_0)$, and $C_a(t_1)$ at the moment when a mutation occurs in $D(r,t_0)$. If a second mutation occurs in the shaded mutation, then the cells located at $a$ and $b$ will be the same.}
\label{figure:diff}
\end{center}
\end{figure}

\vspace{5mm}

If the first mutation occurs in $D(r,t_0)$, then its position $(x_1, y_1)$ is closer to either $a$ or $b$. Without loss of generality, say that $(x_1, y_1)$ is closer to $a$. Again let $r_a$ be the distance between $(x_1, y_1)$ and $a$, and let $C_a(t_1)$ be the cone centered at $a$ with radius $r_a$ at time $t_1$ and radius $0$ at time $(t_1 + \frac{r_a}{c_d})$. A cross-section of this cone is shown in Figure \ref{figure:diff}.

\vspace{3mm}

If the second mutation occurs outside of $C_a(t_1)$, then the first mutation will reach $a$ before interacting with the second mutation. Since the first mutation is outside $V_b(t_0)$, it cannot reach $b$ by time $t_0$, so the sampled cells will be different. 

\vspace{3mm}

If the second mutation occurs inside $C_a(t)$ but outside $M(r,t)$, then the two mutations will interact before the first mutation reaches $a$, meaning that the second mutation will affect $a$. However, the second mutation will not spread to $b$, since it does not start in $V_b(t_0)$. Hence, the cells located at $a$ and $b$ will only be the same if the second mutation occurs in $C_a(t_1) \cap M(r,t_0)$. 

\vspace{5mm}

In summary, we have:
\begin{align}
\label{eq:DabCond}
\P(D_{ab} | E_2) = & \P(X_2 \in C_b(t_1) \triangle C_a(t_1) | X_1 \in M(r,t_0))\P(X_1 \in M(r,t_0)) \\ & + \P(X_2 \notin C_a(t_1) \cap M(r,t_0) | X_1 \in V_a(t_1) \setminus V_b(t_1))\P(X_1 \in V_a(t_1) \setminus V_b(t_1) )\nonumber\\ 
& +\P(X_2 \notin C_b(t_1) \cap M(r,t_0) | X_1 \in V_b(t_1) \setminus V_a(t_1))\P(X_1 \in V_b(t_1) \setminus V_a(t_1))\nonumber
\end{align}

\vspace{5mm}

Since the mutations arise according to a homogenenous Poisson process, we can use the volume calculations for $M(r,t_0)$ and $V_a(t_1) \setminus V_b(t_1)$ to calculate the following probabilities:
\begin{align*}
&\P(X_1 \in M(r,t_0)) = \mu(|M(r,t_0)|)e^{-\mu(|M(r,t_0)|)} \\
&\P(X_1 \in V_a(t_1) \setminus V_b(t_1)) = \P(X_1 \in V_b(t_1) \setminus V_a(t_1)) = \mu(|V_a(t_1) \setminus V_b(t_1)|)e^{-\mu(|V_a(t_1) \setminus V_b(t_1)|)}
\end{align*}

Similarly to the 1-D case:
\begin{align}
\label{eq:Integral1}
&\P(X_2\in C_b(t) \triangle C_a(t) | X_1 \in M(r,t_0))\\
& = \frac{1}{|M(r,t_0)|}\int_{M(r,t_0)}\frac{|C_b(t) \triangle C_a(t)|}{|V_a(t_0) \cup V_b(t_0) \setminus A(x,y,t)|}dxdydt\nonumber,
\end{align}
where $A(x,y,t)$ is the cone-shaped region inside $V_a(t_0) \cup V_b(t_0)$ that is affected by a mutation at $(x,y,t)$.

\vspace{3mm}

 In addition:
\begin{align}
\label{eq:Integral2}
&\P(X_2\in C_a(t_1) \cap M(r,t_0) | X_1 \in V_a(t_0) \setminus V_b(t_0))\\
& = \frac{1}{|V_a(t_0) \setminus V_b(t_0)|} \int_{V_a(t_0) \setminus V_b(t_0)}\frac{|C_a(t) \cap M(r,t_0)|}{|V_a(t_0) \cup V_b(t_0) \setminus A(x,y,t)|}dxdydt.\nonumber
\end{align}


\vspace{7mm}

We next develop formulas to compute the volumes in the previous two displays.
Note that
\begin{equation*}
|C_b(t_1) \triangle C_a(t_1)| = |C_b(t_1)| + |C_a(t_1)| - 2|C_b(t_1) \cap C_a(t_1)|
\end{equation*}
and that $|C_a(t_1)| = \displaystyle\frac{\pi}{3}r_a^2\left(t_1+\frac{r_a}{c_d}\right)$, since $C_a(t_1)$ is a cone with radius $r_a$ and height $t_1+\frac{r_a}{c_d}$. Similarly, $|C_b(t_1)| = \displaystyle\frac{\pi}{3}r_b^2\left(t_1+\frac{r_b}{c_d}\right)$. 

\vspace{7mm}

We next compute $|C_a(t_1)\cap C_a(t_1)|$.
A cross-section of $C_a(t_1)$ has radius $r_a - (s-t_1)c_d$ at time $s$, and a cross-section of $C_b(t_1)$ has radius $r_b - (s-t_1)c_d$ at time $s$. $C_a(t_1)$ and $C_b(t_1)$ will have a nonempty intersection until $r_a - (s-t_1)c_d + r_b - (s-t_1)c_d = r$, i.e. when $s=\displaystyle\frac{r_a+r_b-r}{2c_d}+t_1$.

\vspace{5mm}

If we denote the area of intersection of the cross-sections of $C_a(t_1)$ and $C_b(t_1)$ at time $s$ by $I(s)$, then 
\begin{equation}
\label{eq:area_CaCb}
|C_b(t_1) \cap C_a(t_1)| = \displaystyle\int_{t_1}^{\frac{r_a+r_b-r}{2c_d}+t_1} I(s)ds.
\end{equation}

\vspace{3mm}

$I(s)$ is calculated by summing the areas of the two circular segments, each of which can be calculated by subtracting the area of a triangle from the area of a wedge of the circle
\begin{align*}
I(s) &= R_a^2(s)\cos^{-1}\left(\displaystyle\frac{d_a(s)}{R_a(s)}\right) - d_a(s)\sqrt{R_a^2(s)-d_a^2(s)} \\
&\quad+ \R_b^2(s)\cos^{-1}\left(\displaystyle\frac{d_b(s)}{R_b(s)}\right) - d_b(s)\sqrt{R_b^2(s)-d_b^2(s)},
\end{align*}
where $R_a(s) = r_a - (s-t_1)c_d$, $R_b(s) =  r_b - (s-t_1)c_d$, $d_a(s) = \displaystyle\frac{r^2-R_b^2(s)+R_a^2(s)}{2r}$, and $d_b = \displaystyle\frac{r^2+R_b^2-R_a^2}{2r}.$

\vspace{7mm}

In order to compute the quantity $|C_a(t)\cap M(r,t)|$ used in equation \ref{eq:Integral2}, we need to first determine when the time cross sections of $C_a(t)$ and $V_b(t)$ have nonempty intersection. This occurs when $r_a- (s-t_1)c_d + c_d(t_0-s) > r$, i.e. when $s< \displaystyle\frac{1}{2}\left(\frac{r_a-r}{c_d}+t_0+t_1\right)$. Hence:
\begin{equation}
\label{eq:area_CaM}
|C_a(t) \cap M(r,t)| = \displaystyle \int_{t_1}^{\frac{1}{2}(\frac{r_a-r}{c_d}+t_0+t_1)} \hat{I}(s)ds,
\end{equation}
where
\begin{align*}
\hat{I}(s) &= R_a^2(s)\cos^{-1}\left(\displaystyle\frac{\hat{d_a}(s)}{R_a(s)}\right) - \hat{d_a}(s)\sqrt{R_a^2(s)-\hat{d_a}^2(s)}\\
&\quad +
 \hat{R_b}^2(s)\cos^{-1}\left(\displaystyle\frac{\hat{d_b}(s)}{\hat{R_b}(s)}\right) - \hat{d_b}(s)\sqrt{\hat{R_b}^2(s)-\hat{d_b}^2(s)}.
\end{align*}
$R_a$ is defined above, $\hat{R_b}(s) = c_d(t_0-s)$, $\hat{d_a}(s) = \displaystyle\frac{r^2-\hat{R_b}(s)^2+R_a^2(s)}{2r}$, and $\hat{d_b}(s) = \displaystyle\frac{r^2+\hat{R_b}^2(s)-R_a^2(s)}{2r}$.

\vspace{5mm}

We finally compute $|(V_a(t_0)\cup V_b(t_0))\setminus A(x,y,t)|$.
In pursuit of this define $U_1$ as the region that is affected by the mutation at $(x_1, y_1,t_1)$, i.e., 
\begin{align*}
U_1 = \{ (x,y,s) : |(x,y)-(x_1,y_1)| \leq c_d(s-t_1), t_1 \leq s \leq t_0 \}.
\end{align*}                                                              
Let $u_1(s)$ be the cross-section of $U_1$ at time $s$, i.e., 
\begin{align*}
u_1(s) = \{ (x,y) : |(x,y) - (x_1,y_1)| \leq c_d(s-t_1) \}.
\end{align*}
Observe that $A(x_1,y_1,t_1)$ is the region inside $V_a(t_0) \cup V_b(t_0)$ that is affected by a mutation at $(x_1,y_1,t_1)$, so $A(x_1,y_1,t_1)=U_1 \cap (V_a(t_0) \cup V_b(t_0))$. This of course implies that 
$$
|(V_a(t_0)\cup V_b(t_0))\setminus A(x_1,y_1,t_1)|=|V_a(t_0)\cup V_b(t_0)|-|A(x_1,y_1,t_1)|,
$$

\vspace{5mm}

It thus remains to find $|A(x_1,y_1,t_1)|$. This will be accomplished by looking at the cross-sections of this set for each fixed time $s$.
Define $v_a(s)$ and $v_b(s)$ as the cross section of $V_a$ and $V_b$, respectively, at time $s$, i.e., 
\begin{align*}
v_a(s) = \{ (x,y) : |(x,y) - a| \leq c_d(t_0-s) \} \\ 
v_b(s) = \{ (x,y) : |(x,y) - b| \leq c_d(t_0-s) \}.
\end{align*}

\vspace{7mm}

If $(x_1,y_1, t_1) \in V_a(t_0) \setminus V_b(t_0)$, then $U_1$ will not intersect $V_b(t_0)$, so in this case $A(x_1,y_1,t_1) = U_1 \cap V_a(t_0)$. In order to compute the volume of this set we look at the area of the cross section for each fixed time point. There are three distinct regions for the behavior of the area of this cross section. In the first section of time the cross section of $U_1$ is contained in the cross section of $V_a$:
\begin{align*}
& u_1(s) \cap v_a(s) = u_1(s)  \\
& \iff u_1(s) \subset v_a(s) \\
& \iff r_a+c_d(s-t_1) < c_d(t_0-s)\\ 
& \iff s<\displaystyle \frac{t_1+t_0}{2} - \frac{r_a}{2c_d}.
\end{align*}
In the final section of time the cross section of $V_a$ is contained in the cross section of $U_1$:
\begin{align*}
& u_1(s) \cap v_a(s) = v_a(s)\\
& \iff v_a(s) \subset u_1(s)\\
& \iff r_a+c_d(t_0-s)< c_d(s-t_1)\\
& \iff s>\displaystyle \frac{t_1+t_0}{2} + \frac{r_a}{2c_d}.
\end{align*}

\vspace{5mm}

When $\displaystyle \frac{t_1+t_0}{2} - \frac{r_a}{2c_d} < s < \displaystyle \frac{t_1+t_0}{2} + \frac{r_a}{2c_d}$ :
\begin{align*}
|u_1(s) \cap v_a(s)| &= R_u2(s)\cos^{-1}\left(\displaystyle\frac{d_u(s)}{R_u(s)}\right) - d_u(s)\sqrt{R_u^2(s)-d_u^2(s)} + \hat{R}_a^2(s)\cos^{-1}\left(\displaystyle\frac{\hat{d}_a(s)}{\hat{R}_a(s)}\right) \\
&\quad- \hat{d}_a(s)\sqrt{\hat{R}_a^2(s)-\hat{d}_a^2(s)},
\end{align*}
where $R_u(s) =  c_d(s-t_1)$, $\hat{R_a}(s) = c_d(t_0-s)$, $d_u(s) = \displaystyle\frac{r_a^2-\hat{R}_a^2(s)+R_u^2(s)}{2r_a}$, and $\hat{d}_a(s) = \displaystyle\frac{r_a^2+\hat{R}_a^2(s)-R_u^2(s)}{2r_a}$.

\vspace{5mm}

Thus for $(x_1,y_1, t_1) \in V_a(t_0) \setminus V_b(t_0)$:
\begin{align}
\label{eq:areaA1}
&|A(x_1,y_1,t_1)|\\
& =
 \displaystyle \int_{t_1}^{\frac{t_1+t_0}{2} - \frac{r_a}{2c_d}} |u_1(s)|ds +  \int_{\frac{t_1+t_0}{2} - \frac{r_a}{2c_d}}^{\frac{t_1+t_0}{2} + \frac{r_a}{2c_d}} |u_1(s) \cap v_a(s)|ds +  \int_{\frac{t_1+t_0}{2} + \frac{r_a}{2c_d}}^{t_0} |v_a(s)|ds\nonumber
\end{align}
$|A(x_1,y_1,t_1)|$ is computed analogously when $(x_1,y_1, t_1) \in V_b(t_0) \setminus V_a(t_0)$.

\vspace{10mm}

It remains to compute $|A(x_1,y_1,t_1)|$ when $(x_1,y_1, t_1) \in V_b(t_0) \cap V_a(t_0)$. First note that if $(x_1,y_1, t_1) \in V_b(t_0) \cap V_a(t_0)$:
\begin{align*} 
& u_1(s) \cap (v_a(s)\cup v_b(s)) = u_1(s) \\
& \iff u_1(s) \subset (v_a(s) \cup v_b(s)) \\
& \iff s<\displaystyle \frac{t_1+t_0}{2} - \frac{\min \{r_a,r_b\}}{2c_d}\doteq s_1.
\end{align*}

Once the cross-sections $v_a(s)$ and $v_b(s)$ are no longer intersecting i.e., when $s>t_0-\frac{r}{2c_d}$, then
\begin{align*}
|u_1(s) \cap (v_a(s)\cup v_b(s))| = |u_1(s)\cap v_a(s)| + |u_1(s)\cap v_b(s)|. 
\end{align*}
The two quantities $|u_1(s)\cap v_a(s)|$ and $|u_1(s)\cap v_b(s)|$ can be calculated as shown above.

\vspace{5mm}

 Then for $\displaystyle \frac{t_1+t_0}{2} - \frac{\min \{r_a,r_b\}}{2c_d} < s < t_0-\frac{r}{2c_d}$:
\begin{align*}
|u_1(s) \cap (v_a(s)\cup v_b(s))| = |u_1(s) \cap v_a(s)|+|u_1(s) \cap v_b(s)|-|u_1(s) \cap v_a(s) \cap v_b(s)|.
\end{align*}

\vspace{5mm}

$|u_1(s) \cap v_a(s)|$ and $|u_1(s) \cap v_b(s)|$ can be calculated as shown above, and the quantity $|u_1(s) \cap v_a(s) \cap v_b(s)|$ can be calculated as shown in \cite{Fewell2006}.

\vspace{3mm}

Thus if $(x_1,y_1,t_1)\in V_a(t_0)\cap V_b(t_0)$, then
\begin{align}
\label{eq:areaA2}
&|A(x_1,y_1,t_1)|\\
&=
\int_{t_1}^{s_1} |u_1(s)|ds+\int_{s_1}^{t_0-r/(2c_d)}|u_1(s)\cap (v_a(s)\cup v_b(s))|ds\nonumber.
\end{align}
With \eqref{eq:areaA1} and \eqref{eq:areaA2} we can compute $|A(x_1,y_1,t_1)|$ for arbitrary $(x_1,y_1,t_1)$. We can then use $|A(x_1,y_1,t_1)|$ with \eqref{eq:area_CaM} and \eqref{eq:area_CaCb} to compute \eqref{eq:Integral1} and \eqref{eq:Integral2}. Finally we use \eqref{eq:Integral1} and \eqref{eq:Integral2} compute $P(D_{ab}|E_2)$ based on \eqref{eq:DabCond}.

\section{$I_2$ Calculations}\label{sec:I2calc}
In this section we describe how to compute $I_2(r,t)$ and $I_2(r,\tau)$. First recall from Section \ref{sec:I2} that $R$ is the radius of the clone, $Y$, chosen according to a size-biased pick, and $X$ is the distance of $p$ (a point selected at random from $Y$) from the center of $Y$.

We first describe how to estimate $I_2(r,t)$ based on \eqref{eq:phi_I2}.
In particular, we can generate $K$ i.i.d copies of the vector $(X,R)$, denoted by $\{(X_i,R_i)\}_{i=1}^K$.
Our method for generating $(X_1,R_1)$ based on the time interval $[0,t]$ is as follows. First generate the arrival times of mutations based on a Poisson process with rate $Nu_1s$, denote this set of times by $t_1,\ldots, t_n$.. Then for each mutation calculate the size of its family at time $t$ using the formula \eqref{eq:family_size}, this gives us the collection of family sizes $Y_{1,1},\ldots, Y_{1,n}$ of clones $C_{1,1},\ldots, C_{1,n}$ . Choose a clone $C=C_{[1]}$ via a size biased pick from the collection $C_{1,1},\ldots, C_{1,n}$, and set $R$ to be the radius of $C$. Let $U$ be a uniform random variable on $[0,1]$ independent of $R$ and set $X=R\sqrt{U}$.
With these samples form the estimator
$$
\hat{I}_2(r,t)=\frac{1}{K}\sum_{i=1}^{K}P(p_2\in C|R_i,X_i).
$$

We can also derive an alternative representation for $P(p_2\in C)$ that is more suitable for mathematical analysis.
Denote the conditional density of $X$ given $R=y$ by $f_X(x|R=y)$ and the density of $R$ by $f_R$.
It's easy to see that $f_X(x|R=y)=\frac{2x}{y^2}$ for $x\in (0,y)$ and $0$ otherwise, and therefore
\begin{align}
\label{eq:I2_doubleIntegral}
P(p_2\in C)&=\int_r^{\infty}\int_0^{y-r}\frac{2x}{y^2}f_R(y)dxdy+ \int_{r/2}^r\int_{r-y}^{y}\phi(x,y)\frac{2x}{y^2}f_R(y)dxdy\nonumber\\
&\quad+
\int_r^{\infty}\int_{y-r}^{y}\phi(x,y)\frac{2x}{y^2}f_R(y)dxdy\nonumber\\
&=
\int_r^{\infty}\frac{(y-r)^2}{y^2}f_R(y)dy+ \int_{r/2}^r\int_{r-y}^{y}\phi(x,y)\frac{2x}{y^2}f_R(y)dxdy\nonumber\\
&\quad+
\int_r^{\infty}\int_{y-r}^{y}\phi(x,y)\frac{2x}{y^2}f_R(y)dxdy.
\end{align}
Define
$$
\psi_r(R)=\frac{2}{R^2}\int_{|R-r|}^Rx\phi(x,R)dx
$$
and 
$$
\Phi_r(R)=\begin{cases}\frac{(R-r)^2}{R^2}+\psi_r(R),&\enskip R\geq r\\
\psi_r(R),&\enskip R\in (r/2,r)\\
0,&\enskip R\leq r/2.
\end{cases}
$$
Therefore we see from \eqref{eq:I2_doubleIntegral} that we have $\P(p_2\in C) = \E[\Phi_r(R)]$. 

The formula $\P(p_2\in C) = \E[\Phi_r(R)]$ is difficult to work with due to the complex distribution of $R$. However, an interesting observation is that the distribution of $R$ becomes much simpler if we assume that the sampling occurs at the random detection time $\tau$. In this case define $R(\tau)$ to be the radius of the clone that we choose at time $\tau$, then we can use equation (9) in \cite{foo2014multifocality} to see that conditional on $\tau=t$, $R(\tau)$ has density
$$
f(x|t)=\frac{\mu\gamma_d x^d}{c_d(1-e^{-\theta t^{d+1}})}\exp\left[-\frac{\mu\gamma_dr^{d+1}}{c_d(d+1)}\right]
$$
for $x\leq c_dt$ and zero otherwise. In the conditional density above
$\theta = \mu\gamma c_d^d/(d+1)$. In order to describe the distribution of $R(\tau)$ we then need the distribution of $\tau$, which we can get from (4) of \cite{foo2014multifocality}. In particular define
$$
\phi(t)=\frac{1}{t}\int_0^t\exp\left(-\theta r^{d+1}\right)dr,
$$ 
and $\lambda=Nu_1s$ then $\tau$ has density
$$
f_\tau(t)=\lambda e^{t\lambda(\phi(t)-1)}\left(1-e^{-\theta t^{d+1}}\right).
$$
Therefore we can calculate that
\begin{align*}
P\left(R(\tau)>z\right)&=\frac{\mu\gamma_d\lambda}{c_d}\int_{z/c_d}^{\infty}\int_z^{c_dt}r^d\exp\left[-\frac{\mu\gamma_dr^{d+1}}{c_d(d+1)}\right]dre^{t\lambda(\phi(t)-1)}dt\\
&=
\lambda\int_{z/c_d}^{\infty}\left(\exp\left(-\theta(z/c_d)^{d+1}\right)-\exp\left(-\theta t^{d+1}\right)\right)e^{t\lambda(\phi(t)-1)}dt\\
&=
\exp\left[\lambda (z/c_d)(\phi(z/c_d)-1)\right]-\lambda\left[1-\exp\left(-\theta(z/c_d)^{d+1}\right)\right]\int_{z/c_d}^{\infty}e^{\lambda t(\phi(t)-1)}dt.
\end{align*}
Furthermore we can take derivatives to find that $R(\tau)$ has density given by
$$
\hat f_R(z)=\frac{\lambda\theta (d+1)}{c_d^{d+1}}z^d\exp\left[-\theta\left(\frac{z}{c_d}\right)^{d+1}\right]\int_{z/c_d}^{\infty}e^{t\lambda(\phi(t)-1)}dt.
$$
Note that the density $\hat{f}$ is very similar to the Weibull density, and thus we can generate samples from $\hat{f}$ by using the acceptance rejection algorithm with a proposal distribution based on the Weibull distribution. With these samples from the density $\hat{f}$ then we can use the function $\Phi_r$ to estimate $I_2(r,\tau)$.

Note that when approximating $I_2(r,\tau)$ it is not necessary to simulate the mesoscopic model. We simply generate random variables according to the density $\hat{f}_R$ and then evaluate the function $\Phi_r(R)$. However, when approximating $I_2(r,t)$ it is necessary to simulate the mesoscopic model and thus a greater computational burden.

\end{document}